\documentclass{aa}  

\usepackage{graphicx}
\usepackage{mathtools}
\usepackage{txfonts}
\usepackage{tabularx}
\usepackage{chemformula}
\usepackage{multirow}
\usepackage{booktabs}
\usepackage[]{hyperref}
\hypersetup{
    colorlinks = true,
    linkcolor = {blue}, 
    citecolor = {blue},
    urlcolor = {blue}
}

\newcommand{\per}{{\small $2.72403 ^{+2.34875 \times 10^{-5}} _{-2.28439 \times 10^{-5}}$}}
\newcommand{\Ttra}{{\small $2459691.99424 ^{+0.00017} _{-0.00017}$}}
\newcommand{\rprs}{{\small $0.06903 ^{+0.00020} _{-0.00020}$}}
\newcommand{\bb}{{\small $0.433 ^{+0.013} _{-0.012}$}}
\newcommand{\bbdot}{{\small $-1.59 ^{+38.77} _{-39.35} \times 10^{-6}$}}
\newcommand{\ar}{{\small $4.583 ^{+0.028} _{-0.030}$}}
\newcommand{\lamp}{{\small $91.58 ^{+1.14} _{-1.14}$}}
\newcommand{\lampdot}{{\small $-0.0026 ^{+0.0058} _{-0.0043}$}}
\newcommand{\ECowan}{{\small $203.4 ^{+16.2} _{-16.3}$}}
\newcommand{\COneCowan}{{\small $137.8 ^{+17.0} _{-17.1}$}}
\newcommand{\DOneCowan}{{\small $-23.6 ^{+16.4} _{-16.6}$}}
\newcommand{\CTwoCowan}{{\small 0}}
\newcommand{\DTwoCowan}{{\small 0}}
\newcommand{\Mstar}{{\small $2.054 ^{+0.089} _{-0.087}$}}
\newcommand{\Rstar}{{\small $2.369 ^{+0.023} _{-0.024}$}}
\newcommand{\Tpole}{{\small $7999.24 ^{+75.52} _{-74.96}$}}
\newcommand{\vsini}{{\small $91.70 ^{+2.15} _{-2.23}$}}
\newcommand{\phistar}{{\small $21.58 ^{+1.43} _{-1.54}$}}
\newcommand{\gdc}{{\small 0.22}}
\newcommand{\qone}{{\small $0.177 ^{+0.029} _{-0.027}$}}
\newcommand{\qtwo}{{\small $0.307 ^{+0.091} _{-0.081}$}}
\newcommand{\istar}{{\small $68.42 ^{+1.54} _{-1.43}$}}
\newcommand{\fstar}{{\small $2.86 ^{+0.18} _{-0.17}$}}
\newcommand{\Pstar}{{\small $1.22 ^{+0.03} _{-0.03}$}}
\newcommand{\uone}{{\small $0.258 ^{+0.055} _{-0.056}$}}
\newcommand{\utwo}{{\small $0.162 ^{+0.086} _{-0.083}$}}
\newcommand{\PSI}{{\small $89.46 ^{+1.08} _{-1.08}$}}
\newcommand{\iplanet}{{\small $84.58 ^{+0.19} _{-0.20}$}}
\newcommand{\FdayFs}{{\small $203.4 ^{+16.2} _{-16.3}$}}
\newcommand{\FnFs}{{\small $-71.8 ^{+36.4} _{-36.0}$}}
\newcommand{\phioff}{{\small $-9.78 ^{+6.81} _{-6.82}$}}
\newcommand{\AB}{{\small $0.35 ^{+0.06} _{-0.07}$}}
\newcommand{\ABNight}{{\small $0.19 ^{+0.07} _{-0.07}$}}
\newcommand{\ABRange}{{\small 0.19--0.35}}
\newcommand{\ABthreelow}{{\small $-$0.02}}
\newcommand{\ABthreeup}{{\small 0.53}}
\newcommand{\epsheat}{{\small $0.09 ^{+0.06} _{-0.04}$}}
\newcommand{\epsheatNight}{{\small $0.41 ^{+0.06} _{-0.05}$}}
\newcommand{\epsRange}{{\small 0.09--0.41}}
\newcommand{\epsthreelow}{{\small $-$0.03}}
\newcommand{\epsthreeup}{{\small 0.59}}
\newcommand{\Tday}{{\small $2746 ^{+40} _{-45}$}}
\newcommand{\Tnight}{{\small $1529 ^{+222} _{-209}$}}
\newcommand{\psdpeak}{2.74}
\newcommand{\GPundamped}{{\small $0.55^{+0.36} _{-0.27}$}}
\newcommand{\GPS}{{\small $-20.05 ^{+0.03} _{-0.07}$}}
\newcommand{\GPQ}{{\small $-1.47 ^{+0.54} _{-0.67}$}}
\newcommand{\GPdamped}{{\small $2.19^{+1.35} _{-0.44}$}}
\newcommand{\GPdampedhr}{{\small $52.63^{+32.48}_{-10.48}$}}
\newcommand{\phometricprecCHEOPS}{{\small 12.76~ppm}}
\newcommand{\phometricprecTESS}{{\small 47.59~ppm}}
\newcommand{\RpJup}{{\small $1.591 ^{+0.015} _{-0.016}$}}
\newcommand{\asemimajor}{{\small $0.0505 ^{+0.0005} _{-0.0006}$}}

\newcommand{\boldchange}[1]{#1}
\newcommand{\boldchangenew}[1]{#1}

\begin{document}

   \title{TESS phase curve of ultra-hot Jupiter WASP-189\,b\thanks{The raw and detrended photometric data from TESS, along with codes to generate figures, can be found at: \url{https://github.com/Jayshil/w189Figs}.}}

   \author{
         J.~A.~Patel\inst{1}\thanks{\email{jayshil.patel@astro.su.se}}$^{\href{https://orcid.org/0000-0001-5644-6624}{\includegraphics[scale=0.5]{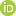}}}$,
         D.~Kitzmann\inst{2,3}$^{\href{https://orcid.org/0000-0003-4269-3311}{\includegraphics[scale=0.5]{Figures/orcid.jpg}}}$,
         A.~Brandeker\inst{1},
         T.~G.~Wilson\inst{4},
         A.~Deline\inst{5},
         M.~Lendl\inst{5}, and
         V.~Singh\inst{6}
        }
   \authorrunning{Patel et al.}

   \institute{
             \label{inst:1} Department of Astronomy, Stockholm University, AlbaNova University Center, 10691 Stockholm, Sweden \and
             \label{inst:2} Weltraumforschung und Planetologie, Physikalisches Institut, Universität Bern, Gesellschaftsstrasse 6, 3012 Bern, Switzerland \and
             \label{inst:3} Center for Space and Habitability, Universität Bern, Gesellschaftsstrasse 6, 3012 Bern, Switzerland \and
             \label{inst:4} Department of Physics, University of Warwick, Gibbet Hill Road, Coventry CV4 7AL, UK \and
             \label{inst:5} Observatoire astronomique de l'Université de Genève, Chemin Pegasi 51, 1290 Versoix, Switzerland \and
             \label{inst:6} INAF, Osservatorio Astrofisico di Catania, Via S. Sofia 78, 95123 Catania, Italy
             }

   \date{Received; accepted}

  \abstract
   {The thermal structures of highly irradiated ultra-hot Jupiters can deviate substantially from those of cooler hot Jupiters. For planets orbiting rapidly rotating, and consequently oblate, host stars, photometric light curves provide a unique opportunity to measure the spin–orbit angle. Moreover, in systems with significant spin–orbit misalignment, the stellar oblateness can induce observable orbital precession.
   }
   {We wish to study the atmosphere and orbital architecture of an ultra-hot Jupiter WASP-189\,b, orbiting around a hot A-type star.}
   {We use the photometric phase curves and gravity-darkened transits of WASP-189\,b observed with the Transiting Exoplanet Survey Satellite (TESS), complemented with the archival observations from CHaracterising ExOPlanet Satellite (CHEOPS).}
   {We detected a phase curve signal with significant occultation depth of \ECowan~ppm, while the nightside flux, \FnFs~ppm, is consistent with zero at 2$\sigma$. We invert the phase curve signal to construct the temperature map of the planet. The map was subsequently used to estimate the Bond albedo and heat redistribution efficiency, the expected median ranges of which are found to be \ABRange\, and \epsRange, respectively. Finally, we analysed gravity-darkened transits to find that the planet is in polar orbit with the spin-orbit angle of \PSI~deg. We found no hint of orbital precession while comparing our results with those from the literature.}
   {Our observations, together with atmospheric modelling, suggest that the dayside emission of WASP-189\,b in TESS and CHEOPS bandpasses is dominated by thermal emission from an atmosphere with extremely inefficient heat transport and negligible contribution from reflected light.}


   \maketitle
%
%-------------------------------------------------------------------

\section{Introduction}\label{sec:intro}

The population of hot Jupiters with extremely high temperatures (T$_{\text{eq}}$~$\gtrsim$~2200~K) has emerged as a distinct class of planets, known as ultra-hot Jupiters. They have dissimilar atmospheric properties compared to the cooler hot Jupiters. In particular, while the cooler hot Jupiters have absorption features in their occultation depth spectra \citep[e.g.,][]{2014Sci...346..838S}, the ultra-hot Jupiters' spectra are either featureless blackbody-like or have emission features \citep[e.g.,][]{2017ApJ...850L..32S, 2018AJ....156...10M, 2023Natur.620..292C}. The change from absorption to emission features is a result of changing the temperature profile from non-inverted to inverted.
\citet{2021NatAs...5.1224M} have discovered the trend of spectral features in occultation depth spectra of hot Jupiters: they found that the spectral features change from absorption to emission and finally become muted with increasing equilibrium temperature. The emergence of short-wave absorbers, such as TiO and VO, in the upper atmosphere at higher temperatures ($\gtrsim$~2000~K), creates an inverted temperature profile that results in emission features in the spectra \citep[e.g.,][]{2021NatAs...5.1224M}. At even higher temperatures ($\gtrsim$~3000~K), although the temperature profile is still inverted, molecular dissociation and the presence of H$^-$ continuum opacity mute the spectral features \citep[e.g.,][]{2018ApJ...855L..30A, 2018A&A...617A.110P, 2021NatAs...5.1224M}. The H$^-$ opacity also causes a low dayside geometric albedo of the planet.

Hot Jupiters have a strong day-night temperature difference \citep[e.g., see][for a review]{2015AREPS..43..509H}. The temperature contrast is expected to increase with temperature because the radiative cooling becomes more efficient compared to the efficiency of the heat transport to the nightside of the planet \citep[e.g.,][]{2016ApJ...821...16K}. However, this trend might break for ultra-hot Jupiters: the dissociated molecular H$_2$, which forms atomic H on the dayside, because of high temperatures, would recombine to again create H$_2$ on the nightside. This can increase the day-night energy transport and thus effectively reduce the day-night temperature contrast \citep[][]{2018ApJ...857L..20B}. Phase curve observations of several ultra-hot Jupiters, such as KELT-9\,b, are already showing hints of this in terms of smaller, yet inconsistent with zero, heat redistribution efficiency \citep[e.g.,][]{2020ApJ...888L..15M, 2022A&A...666A.118J}.

WASP-189\,b, an ultra-hot Jupiter around a hot, fast-rotating A-type star \citep[][]{2018arXiv180904897A} with an equilibrium temperature of \boldchange{3353$^{+27}_{-34}$~K} \citep[][]{2020A&A...643A..94L}, is one of the prime targets to study the dayside emission and thermal structure of ultra-hot Jupiters. Its high-resolution observations in transmission have successfully detected several species including TiO, V, Fe, Fe$^+$, Ti, Ti$^+$, Cr, Mg, Mg$^+$, Mn, H, Na, Ca, Ca$^+$, Ni, Sr, Sr$^+$, Ba$^+$, \citep[][]{2022NatAs...6..449P, 2023A&A...678A.182P, 2023ApJ...954L..23S, 2024A&A...685A..60P}. At the same time, the high-resolution dayside emission spectroscopy from the ground further identified CO, Fe \citep[][]{2022A&A...661L...6Y, 2024AJ....168..148D, 2025A&A...693A..72L}. The presence of short-wave absorbers, like TiO, results in an inverted temperature profile \citep[][]{2022A&A...661L...6Y, 2025A&A...693A..72L}, which is previously seen for other ultra-hot Jupiters \citep[][]{2021NatAs...5.1224M}. The non-detection of Fe in near-infrared high-resolution transmission spectroscopy by \citet{2025A&A...700A...9V} was thought to be because of the presence of H$^-$. CHaracterising ExOPlanet Satellite \citep[CHEOPS;][]{2021ExA....51..109B} has observed several occultations and phase curves of WASP-189\,b with its optical bandpass \citep[][]{2020A&A...643A..94L, 2022A&A...659A..74D}. They detected a significant occultation depth in the optical bandpass of CHEOPS, which, they suggested, could be explained solely by thermal emission from the planet in the case of extremely inefficient heat redistribution. \citet{2022A&A...659A..74D} suggested that the reflective light might contribute to the occultation depth observed in the CHEOPS bandpass, and places an upper limit of 0.48 on the geometric albedo. The ultra-precise transit observations with CHEOPS further showed gravity darkening that subsequently led \citet{2020A&A...643A..94L, 2022A&A...659A..74D} to determine its orbital architecture. The planet was found to be in a polar orbit around its host star.

It is difficult to constrain the reflective properties of the dayside with only one CHEOPS photometric occultation depth. We here present another photometric observation of WASP-189\,b taken with the Transiting Exoplanet Survey Satellite \citep[TESS;][]{2014SPIE.9143E..20R}. While CHEOPS and TESS bandpasses overlap for a range of wavelengths, the TESS bandpass is more sensitive to longer wavelengths. In the absence of other spectroscopic observations of WASP-189\,b, combined TESS and CHEOPS observations can constrain the thermal properties of the planetary atmosphere, which is the prime goal of the present work. Furthermore, since the TESS observations are obtained almost two years after the CHEOPS observations, we also search for the orbital precession, which can happen for planets in misaligned orbits around fast-rotating oblate stars. Sect.~\ref{sec:obs_data_analysis} describes our manual data reduction procedure to extract photometry from the TESS Target Pixel Files (TPFs), and details the planetary, stellar, and instrumental models that we used to fit the data. Sect.~\ref{sec:res} presents the results and discusses their meaning for the planet and its atmosphere. Finally, we discuss stellar variability in Sect.~\ref{sec:stel_var} followed by a conclusion in Sect.~\ref{sec:conclusions}.

%--------------------------------------------------------------------
\section{Observations and data analysis}\label{sec:obs_data_analysis}

\subsection{Observations and data reduction}\label{subsec:obs}

WASP-189 was observed with TESS during its extended mission in Sector 51. The raw data were reduced by the TESS Science Processing Operation Center \citep[SPOC;][]{2016SPIE.9913E..3EJ} to produce PDC-SAP photometry \citep[Presearch Data Conditioning Simple Aperture Photometry;][]{2012PASP..124.1000S, 2014PASP..126..100S}. The PDC-SAP light curve, shown in Fig.~\ref{fig:pdc_scal_comp}, has large gaps in it. In the hope of recovering some of the data, we decided to use our own data reduction pipeline to reduce TESS photometry from TPFs. We first computed a simple aperture photometry by summing up all counts inside the aperture, where the aperture is defined by selecting 15 brightest pixels in the median 2D TPF. We subtract the median of the background flux from all pixels inside the aperture before calculating the photometry. We removed all points with NaN values and 20$\sigma$ outliers. The resultant photometry, shown in Fig.~\ref{fig:pdc_scal_comp}, is heavily affected by the instrumental systematics. We employ the pixel-level decorrelation technique that performs principal component analysis (PCA) on pixel-level light curves of all pixels within the aperture to model instrumental systematics. The first few principal components (PCs) should trace the instrumental systematics. To select how many PCs to include in our model, we performed a linear detrending of simple aperture photometry with model generated from a set of PCs. We added one by one PCs to the model and kept those PCs that produce a detrended photometry with the lowest median absolute deviation (MAD). We found that the use of the first four PCs minimised the MAD, which we eventually used as linear decorrelation vectors in our light curve analysis to model the instrumental systematics (see, Sect.~\ref{subsec:misc_post}). 
We masked all points with a high background larger than 100~e$^-$/s. This technique was previously used to model the instrumental systematics of Spitzer \citep{2015ApJ...805..132D}, K2 \citep{2016AJ....152..100L}, and the James Webb Space Telescope \citep[JWST;][]{2024A&A...690A.159P}.

\subsection{Gravity darkened transit}\label{subsec:transit_model}

The host star WASP-189 is a hot (T$_{\text{eff}}$ = 8000 $\pm$ 80 K), A-type star \citep[][]{2020A&A...643A..94L}. As with the other such hot stars, WASP-189 is a rapid rotator with $v\sin{i} = 93.1 \pm 1.7$~km/s \citep{2020A&A...643A..94L}. A centrifugal force generated by the rapid stellar rotation distorts the shape of the star to make it oblate: the stellar equatorial radius will increase compared to its radius at the pole. \citet{1924MNRAS..84..684V} showed that the local surface gravity, and thus the local temperature and resulting brightness, will change as a function of latitude on an oblate star: the stellar equator would appear darker compared to the poles, a phenomenon called gravity darkening. If a planet in a misaligned orbit were to transit a gravity-darkened oblate star, it would block the stellar disk of varying brightness during its passage. This would lead to a peculiar asymmetric transit shape that depends on the exact geometry of the orbit and stellar orientation \citep{2009ApJ...705..683B}. It is therefore possible to extract information regarding the orbital geometry just from observing the transit. Indeed, gravity darkening induced asymmetric transit was observed previously for several planets, which constrained stellar and orbital orientation relative to the observer \citep[][]{2022A&A...658A..75H, 2022A&A...666A.118J}.

\citet{2020A&A...643A..94L} and \citet{2022A&A...659A..74D} have already observed the gravity darkened transit of WASP-189\,b using ultra-high precision photometry from CHEOPS. Following their results, we decided to fit a gravity-darkened transit model to the TESS data. We used the open-source Python package \texttt{PyTransit} \citep{2015MNRAS.450.3233P} to do this\footnote{\url{https://pytransit.readthedocs.io/en/latest/}}. \texttt{PyTransit} first generates a discrete cell grid on the stellar surface and calculates the local surface gravity on these cells by assuming a star whose outer layers are deformed due to rapid rotation. \texttt{PyTransit} then uses von Zeipel's theorem \citep{1924MNRAS..84..684V} to compute the surface temperature at each discrete cell on the stellar surface. A \texttt{PHOENIX} synthetic stellar spectrum \citep{2013A&A...553A...6H}, the TESS instrument response function, and the quadratic limb darkening law were used to compute the flux emitted from each discrete cell. \texttt{PyTransit} finally uses orbital parameters to estimate the light blocked by the planet during transit.

The gravity darkened transit model in \texttt{PyTransit} is parametrised by the planet-to-star radius ratio ($R_p/R_\star$), the stellar density ($\rho_\star$), the stellar rotation period ($P_\star$), the stellar obliquity with respect to the \boldchange{plane} of the sky ($\phi_\star = \pi/2 - i_\star$, where $i_\star$ is the stellar inclination), the gravity darkening parameter ($\beta$), the limb darkening coefficients ($u_1$ and $u_2$ for the quadratic law), the transit time ($t_0$), the planetary orbital period ($P$), the scaled semi-major axis ($a/R_\star$), the orbital inclination ($i_p$), the projected orbital obliquity ($\lambda_p$), the eccentricity ($e$) and the argument of periastron passage ($\omega$). We directly fitted for all these parameters except $\rho_\star$, $P_\star$, ($u_1$, $u_2$), and $i_p$. We computed $P_\star$ from $\phi_\star$, projected rotation speed ($v\sin{i_\star}$), and radius ($R_\star$), which we treated as free parameters in our analysis, as:

\begin{equation}\label{eq:pstar}
    P_\star = \frac{2\pi R_\star}{v\sin{i_\star}} \sin{i_\star} = \frac{2\pi R_\star}{v\sin{i_\star}} \sin{(\pi/2-\phi_\star)}.
\end{equation}

Furthermore, we can estimate the density of an oblate star using stellar oblateness ($f_\star = 1 - R_{\text{pole}}/R_\star$, where $R_{\text{pole}}$ is the pole radius), stellar mass, and rotation period \citep[e.g.,][]{2022A&A...659A..74D},

\begin{equation}\label{eq:fstar}
    f_\star = \frac{1}{1 + \frac{GM_\star P_\star^2}{2\pi^2 R_\star^3}} = \frac{3\pi}{2G\rho_\star P_\star^2}.
\end{equation}

\noindent \boldchangenew{The second equality in above equation is obtained by using the definition of stellar density for an oblate star,}

\begin{equation}\label{eq:stellar_density}
    \rho_\star = \frac{M_\star}{\frac{4}{3}\pi\cdot(R_{\text{equator}} \cdot R_{\text{equator}} \cdot R_{\text{pole}})} = \frac{M_\star}{\frac{4}{3}\pi R_\star^3(1-f_\star)},
\end{equation}

\noindent in the second term of Eq.~\ref{eq:fstar}. Here, $R_{\text{equator}}$ is the stellar radius at the equator, which we take as $R_\star$, and $R_{\text{pole}} = R_\star(1-f_\star)$.

We directly fitted for $M_\star$ to obtain the stellar density from $P_\star$ and $R_\star$ using Eqs.~\ref{eq:pstar} and \ref{eq:fstar}. $i_p$ and $i_\star$ are computed from $b$ (impact parameter) and $\phi_\star$, respectively.

Instead of directly fitting for quadratic limb-darkening coefficients ($u_1$ and $u_2$), we used \citet{2013MNRAS.435.2152K} parametrisation to fit for limb darkening coefficients. Finally, we fitted for $b$ in our analysis instead of fitting for $i_p$.

Many of the planetary orbital parameters used by \texttt{PyTransit} to generate the gravity-darkened transit model are highly correlated. Fitting for all of them without any prior knowledge can result in degeneracies among model parameters \citep[e.g.,][]{2022A&A...658A..75H}. It is generally recommended to put informative priors on at least some of these parameters based on either Doppler tomography or other previous observations \citep[][]{2015ApJ...805...28M, 2022A&A...658A..75H}. \citet{2022A&A...659A..74D} recently used CHEOPS transit observations to constrain planetary and stellar parameters for the WASP-189 system. Considering this, and given that the photometric precision of our 5 transits is lower than CHEOPS, we decided to put informative priors on most of the planetary and stellar parameters based on the analysis by \citet{2022A&A...659A..74D}. A notable exception to this are the limb darkening coefficients, which could be wavelength dependent. We thus decided to put uninformative priors on their \citet{2013MNRAS.435.2152K} parametrisation \citep[e.g., see,][]{2022AJ....163..228P}.
The full list of parameters and priors used on them is provided in Table~\ref{tab:fitted_transit_param}. Since we use normal priors based on previous CHEOPS observations, our derived parameters will be an update on their literature values, without the need for simultaneous analyses of CHEOPS and TESS data.

\subsection{Occultation and orbital phase curve}\label{subsec:pc_model}

As demonstrated by \citet{2022A&A...659A..74D}, the shape of the occultation light curve for an oblate star would be slightly different from that for a spherical star. Following their method, we set limb-darkening and gravity darkening coefficients in the \texttt{PyTransit} gravity darkened transit model to zero to obtain the occultation light curve. In doing so, we also set the ``transit'' time to the occultation time, $\lambda'_p = -\lambda_p$, and $\omega' = \omega + \pi$. In the end, we normalise the occultation signal such that the flux inside the occultation is set to zero and the out-of-occultation flux is one. Normalisation was done so that we can directly multiply our phase curve model with the normalised occultation model to obtain flux variations due to the planet.

We model the out-of-transit/occultation variation using the implementation of \citet{2008ApJ...678L.129C} phase curve model by \citet{2023Natur.620...67K} in our analysis, which is defined as

\begin{equation}\label{eq:cowan_agol_08}
    \begin{split}
        F_p = E &+ C_1 (\cos{\omega_p t_e} - 1) + D_1\sin{\omega_p t_e} \\
            &+ C_2 (\cos{2\omega_p t_e} - 1) + D_2\sin{2\omega_p t_e}.
    \end{split}
\end{equation}

\noindent Here, $E$, $C_1$, $D_1$, $C_2$ and $D_2$ are phase curve coefficients, $t_e$ is the time since mid-occultation and $\omega_p = 2\pi/P$. Eq.~\ref{eq:cowan_agol_08} is defined such that the parameter E gives the occultation depth. Being a simple sinusoidal function, this function is fast to evaluate. We note here that the above phase curve model only contains the planetary model. We do not model the ellipsoidal variations and the Doppler beaming since their amplitudes are expected to be very small \citep[on the order of 10 and 1 ppm, respectively;][]{2022A&A...659A..74D}. It is possible to invert the \citet{2008ApJ...678L.129C} phase curve function to obtain the temperature map of the planet.

We fitted both the first-order (i.e., setting $C_2=D_2=0$ in Eq.~\ref{eq:cowan_agol_08}) and second-order (non-zero $C_2$ and $D_2$) phase curve models to the data and compared the Bayesian evidence. Upon finding that the first-order model was heavily favoured compared to the second-order model, we fixed $C_2$ and $D_2$ to zero in our final analysis. We put wide, uninformative priors on phase curve parameters in our data analysis (see Table~\ref{tab:fitted_transit_param}).

\subsection{Light travel time delay, noise sources, and posterior sampling}\label{subsec:misc_post}

The light signal from the planet and the star does not reach the observer simultaneously because of the finite light travel time. We have implemented a correction of the light travel time in our analysis, assuming a circular orbit and following the method from \citet{2022A&A...659A..74D}:

\begin{equation}
    t_{\text{ref}} = t_{\text{obs}} - \frac{a}{c} \left[ 1 - \cos{\left( 2\pi \frac{t_{\text{obs}}-t_0}{P} \right)} \right]\sin{i_p}.
\end{equation}

\noindent Here, $a$ is the semi-major axis and $c$ is the speed of light. We essentially computed our planetary and stellar models at $t_{\text{ref}}$ for observed $t_{\text{obs}}$ while simultaneously fitting for $a$, $t_0$, $P$ and $i_p$.

\begin{figure*}
    \centering
    \includegraphics[width=\linewidth]{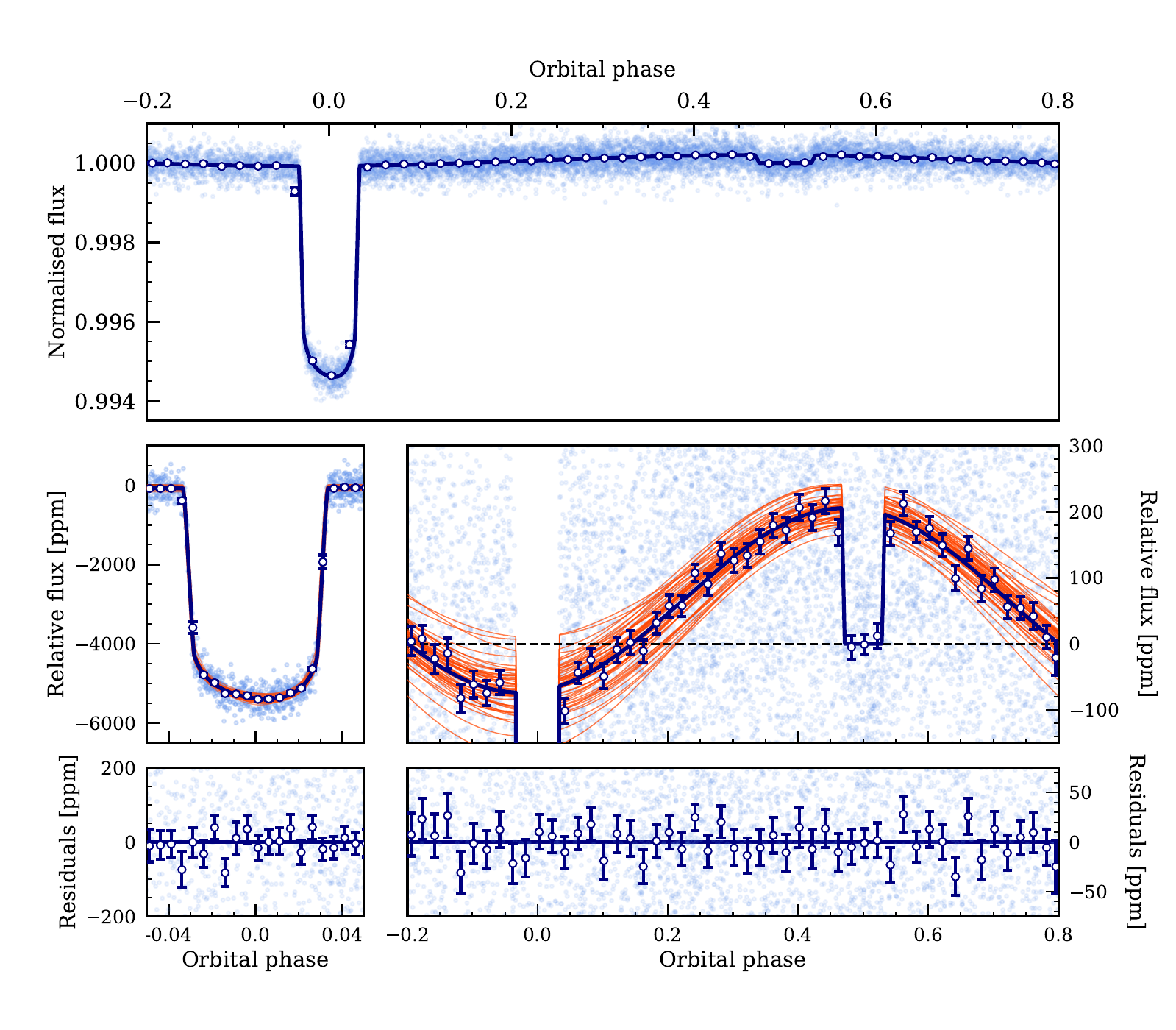}
    \caption{Detrended and phase-folded TESS data along with the best-fitted model. The light and dark blue points show the unbinned and binned data points. The dark blue and orange lines are the median model and models computed from randomly selected posteriors, respectively. A full phase curve with a transit and occultation is shown in the top panel, while the middle panel shows a zoom-in on the transit (middle left) and occultation/phase curve (middle right). The dashed line in the middle right panel represents the level of stellar flux (i.e., flux level during the occultation). The gravity darkened asymmetric transit and a phase variation, along with an occultation, are clearly visible in the middle panel. The bottom panel shows the residuals after subtracting the median model from the raw data.}
    \label{fig:data_models}
\end{figure*}

The simple aperture photometry has strong instrumental systematics as described in Sect.~\ref{subsec:obs}. We included a linear model in the first four PCs from our PCA to model the instrumental systematics. \boldchange{We also tested if our results are sensitive to the number of PCs used in the analysis. While transit parameters are insensitive to the number of PCs, phase curve parameters can change if we include only one or two PCs, as very few PCs cannot model the instrumental systematics properly. This also results in large uncertainties on phase curve parameters. However, after inclusion of the third and more PCs, the values of phase curve parameters stabilise to the values we report here.}
While \boldchange{the detrending using PCs} should take care of most of the instrumental noise, we found an additional noise source that could either be astrophysical or leftover instrumental noise (see Sect.~\ref{sec:stel_var} for more details). We fit this noise with the help of a Gaussian process (GP) model based on the Simple Harmonic Oscillator (SHO) kernel as implemented in \texttt{celerite2} \citep{celerite1, celerite2}.

We used nested sampling methods \citep{2004AIPC..735..395S, 10.1214/06-BA127}, in particular, dynamic nested sampling \citep{2019S&C....29..891H}, implemented in the \texttt{dynesty} \citep{2020MNRAS.493.3132S} package to sample from posterior distribution. The median and 16-84th percentile confidence intervals for our fitted parameters are tabulated in Table~\ref{tab:fitted_transit_param}.

%--------------------------------------------------------------------

\section{Results and discussion}\label{sec:res}

\subsection{Planetary orbit and orbital precession}\label{subsec:orbital_arch}

Our results from the gravity-darkened transit analysis, shown in Table~\ref{tab:fitted_transit_param}, corroborate the results from \citet{2022A&A...659A..74D}, which is expected given our choices of priors. Our prior choice also implies that the values for \boldchange{all} planetary and stellar parameters\boldchange{, except for parameters specific to TESS, e.g., limb darkening coefficients and phase curve parameters,} that we derive are updated \boldchange{from previous analyses}. The best-fitted gravity-darkened transit model, along with the data, is shown in Fig.~\ref{fig:data_models}. Using our best-fitted planetary parameters, we constrain the 3D spin-orbit angle \citep{2022A&A...658A..75H, 2022A&A...659A..74D}\boldchange{,}

\begin{equation}
    \Psi = \arccos{\left( \cos{i_\star}\cos{i_p} + \sin{i_\star} \sin{i_p} \cos{\lambda_p} \right)},
\end{equation}

\noindent which defines the 3D angle between the stellar spin axis and the planetary orbital axis. We constrain its value to \PSI~deg, suggesting a polar orbit. This value of $\Psi$ is consistent with the literature values from \citet{2020A&A...643A..94L} and \citet{2022A&A...659A..74D}. Apart from $\Psi$, although we have properly constrained system parameters, there remains a degeneracy among the system parameters $i_\star$, $i_p$ and $\lambda_p$ inherent to a gravity darkening transit model. The issue is that it is not possible to distinguish between ($i_\star$, $i_p$, $\lambda_p$) and ($180^\circ - i_\star$, $180^\circ - i_p$, $-\lambda_p$). 

The polar orbit of WASP-189\,b and the stellar oblateness provide good conditions for planetary orbital precession. The stellar quadruple moment, $J_2$, can perturb the orbit and cause precession \citep{2015ApJ...805...28M}. The precession can be detected as a temporal evolution of $b$ (or, $i_p$) and $\lambda_p$ \citep{2022A&A...658A..75H}. If detected, this would provide good constraints on $J_2$. We performed another analysis of TESS data \boldchange{alone} to see if there is any change in $b$ and $\lambda_p$\boldchange{, compared to the CHEOPS results \citep{2022A&A...659A..74D}}.

\citet{2022A&A...659A..74D} observed two high-precision phase curves of WASP-189\,b that contain three and a half transits. The TESS observations of WASP-189 occurred about two years after the first \citet{2022A&A...659A..74D} CHEOPS observations. This long baseline provides a unique opportunity to check if the orbit of WASP-189\,b is precessing. Our main analysis of TESS data, described in Sect.~\ref{sec:obs_data_analysis}, is highly influenced by the priors since we provided normal priors on $b$ and $\lambda$. We therefore reanalysed the dataset in the same manner as described in Sect.~\ref{sec:obs_data_analysis}; however, this time, we put uniform priors $\mathcal{U}(0.35, 0.50)$ and $\mathcal{U}(86,97)$ on $b$ and $\lambda_p$, respectively (here, $\mathcal{U}(a,b)$ represents uniform priors between $a$ and $b$). We get $b={\small 0.432 ^{+0.016} _{-0.018}}$ and $\lambda_p = {\small 89.88 ^{+3.85} _{-2.70}}$~deg, which are within 1$\sigma$ of their corresponding values from \citet[][]{2022A&A...659A..74D}, $b={\small 0.433 \pm 0.020}$ and $\lambda_p = {\small 91.7 \pm 1.2}$~deg. The rate of change of $b$ and $\lambda_p$\boldchange{, measured for the first time for WASP-189\,b,} are \bbdot~yr$^{-1}$ and \lampdot~deg~yr$^{-1}$, respectively, both of which are consistent with zero at 1$\sigma$. This means that we do not detect any significant orbital precession. However, this is not surprising given that the measured rate of change of $b$ for other hot Jupiters is on the order of 0.01~yr$^{-1}$ \boldchange{\citep[e.g.,][]{2012MNRAS.421L.122S, 2015ApJ...810L..23J, 2022ApJ...931..111S, 2024PASJ...76..374W}}.The rate of change of b is directly proportional to $J_2$ \citep{2022MNRAS.512.4404W}. The exact value of $J_2$ depends on the star: it was measured to be on the order of 10$^{-4}$--10$^{-5}$ for the early to late A-type stars KELT-9 and TOI-1518 \citep{2022ApJ...931..111S, 2024PASJ...76..374W}, so we should expect a similar value of $J_2$ for WASP-189, which is also an A-type star. Therefore, we should expect a similar rate of change of $b$ for WASP-189\,b. However, the rate is also directly proportional to $\cos{\Psi}$ and inversely proportional to a/R$_\star$ and $P$ \citep{2022MNRAS.512.4404W}. WASP-189\,b has slightly larger values of a/R$_\star$ and $P$ compared to KELT-9\,b and TOI-1518\,b, with $\Psi \approx 90^\circ$, thereby decreasing $\cos{\Psi}$. This can decrease the rate of change of $b$ to less than 0.01~yr$^{-1}$. Our current observations do not significantly constrain such a small change. Observations over a long baseline might help in detecting orbital precession.

\subsection{Atmospheric properties}\label{subsec:atmosphere}

We detect a significant phase curve signal in our TESS dataset, which we modelled using a \citet{2008ApJ...678L.129C} phase curve model (see Fig.~\ref{fig:data_models} for the data and the fitted phase curve model). Parametrisation of this model (Eq.~\ref{eq:cowan_agol_08}) allows us to estimate occultation depth and nightside flux as flux at the time of occultation (i.e., $t_e=0$) and transit ($t_e=P/2$), respectively. According to this formulation, the occultation depth and nightside flux would be $E$ and $E-2C_1$, respectively, which we compute to be \ECowan~ppm and \FnFs~ppm. While we constrain the occultation depth at more than 10$\sigma$, the nightside flux is consistent with zero at 2$\sigma$. 
The negative value of the nightside flux, which is consistent with zero at 2$\sigma$, is likely because of random chance, but it might be because of instrumental noise or uncorrected ellipsoidal variations \citep[e.g.,][]{2025A&A...699A.150D}. However, we note that the ellipsoidal variation is expected to be small for the WASP-189 system (see Sect.~\ref{subsec:pc_model}). We find that the phase offset is consistent with zero, \phioff~deg. \boldchange{Our measurement of the TESS occultation depth, together with the archival CHEOPS occultation depth, constrains the thermal structure of the atmosphere for the first time as we describe below.}

\subsubsection{Atmospheric modelling}\label{subsubsec:atmos_model}

\citet{2022A&A...659A..74D} found an occultation depth of {\small $96.5 ^{+4.5} _{-5.0}$~ppm} with their CHEOPS observations. Both CHEOPS and TESS bandpasses may contain thermal and reflected light contributions to their dayside emission. We can uniquely constrain the geometric albedo (A$_g$) and dayside brightness temperature of the planet \boldchange{given the two observational constraints on the occultation depth from TESS and CHEOPS and} assuming a flat A$_g$ spectrum \citep[e.g.,][]{2024A&A...683A...1S}, resulting in {\small A$_g$~=~$0.24 ^{+0.07} _{-0.07}$} \boldchange{and a dayside temperature of {\small $3121^{+145}_{-181}~K$}}. However, the assumption of a flat geometric albedo spectrum may not be correct.

The atmosphere of WASP-189\,b was modelled by \citet{2020A&A...643A..94L} to explain CHEOPS observations of the planet’s dayside. In their study, the HELIOS atmospheric model \citep{Malik2017AJ....153...56M, Malik2019AJ....157..170M} was used to generate a grid of self-consistent atmospheres as a function of the heat redistribution efficiency, $\epsilon$, and an artificial geometric albedo, $A_\mathrm{g}$. The latter was introduced to account for unknown atmospheric scatterers. Furthermore, this artificial albedo was necessary because HELIOS does not calculate the geometric albedo of an atmosphere, as its radiative transfer scheme only provides angular-integrated fluxes for a single atmospheric column.

For the present study, we adopted the atmospheric structures for WASP-189\,b from \citet{2020A&A...643A..94L}, specifically the HELIOS models without the artificial albedo. To obtain occultation spectra that include the contribution of scattered light, we post-processed the resulting temperature–pressure structures with the multi-stream radiative transfer code C-DISORT \citep{Hamre2013AIPC.1531..923H}, the C-version of the discrete ordinate radiative transfer solver DISORT \citep{Stamnes1988ApOpt}. DISORT solves the one-dimensional radiative transfer equation, including both thermal emission and scattering, and provides not only average quantities such as fluxes and mean intensities, but also the full angular distribution of the radiation field.

To calculate the geometric albedo $A_\mathrm{g}$ contributing to the occultation depth, we solved the radiative transfer problem across the illuminated hemisphere, accounting for the local stellar zenith angle at each latitude ($\Theta$) and longitude ($\Phi$). For simplification, we assume that the dayside-average atmospheric structure obtained from HELIOS is representative for all longitudes and latitudes across the dayside hemisphere.
From the DISORT solution, we extracted the reflected and emitted intensities in the direction of the observer. The reflected component was then integrated over the planetary hemisphere to obtain the Sobolev fluxes at zero phase angle
\begin{equation}
    F_0 = \int_{-\pi/2}^{\pi/2} \int_0^{\pi/2} \rho(\mu, \mu_*, \Theta, \Phi) \, \mu \mu_* \, \mathrm d \Theta \, \mathrm d \Phi,
\end{equation}
where $\mu$ the cosine of the polar angle, $\mu_*(\Theta, \Phi)$ the cosine of the local stellar zenith angle, and $\rho$ the reflection coefficient (see \citet{Heng2021NatAs...5.1001H}). The reflection coefficient is defined as
\begin{equation}
    \rho(\mu, \mu_*, \Theta, \Phi) = \frac{I_0(\mu, \Theta, \Phi)}{I_* \mu_*(\Theta, \Phi)}
\end{equation}
where it relates the incident stellar intensity $I_\ast$ to the intensity reflected towards the observer $I_0$. Using the Sobolev fluxes, the geometric albedo was then calculated as
\begin{equation}
    A_\mathrm{g} = \frac{2 F_0}{\pi} \ ,
\end{equation}
see \citet{Sobolev1975lpsa.book.....S} for details. The corresponding occultation depth is given by
\begin{equation}
  \frac{F_{\mathrm{p}}}{F_{\mathrm{*}}} = A_{\mathrm{g}} \left(\frac{R_\mathrm{p}}{a}\right)^2 + \frac{F_{\mathrm{p}}^{\mathrm{therm}}}{F_{\mathrm{*}}} \left(\frac{R_\mathrm{p}}{R_*}\right)^2 \ ,
\end{equation}
where $F_{\ast}$ is the stellar flux and $F^{\mathrm{therm}}_{\mathrm p}$ is the planet’s thermal emission.

The atmospheric chemical composition was modelled using the open-source code \textsc{FastChem}\footnote{\url{https://github.com/NewStrangeWorlds/FastChem}} \citep{Stock2018MNRAS.479..865S, Stock2022MNRAS.517.4070S, Kitzmann2024MNRAS.527.7263K}, assuming chemical equilibrium. Given the extreme dayside temperatures of WASP-189\,b, this assumption is well justified \citep{Kitzmann2018ApJ...863..183K}. Although condensates were included in the calculations, none were found to be thermally stable under these conditions. Consequently, in our model calculations, the only source of scattering on the dayside is Rayleigh scattering by \ch{H2}, He, and H.

\begin{figure}%[ht]
    \centering
    \includegraphics[width=0.9\linewidth]{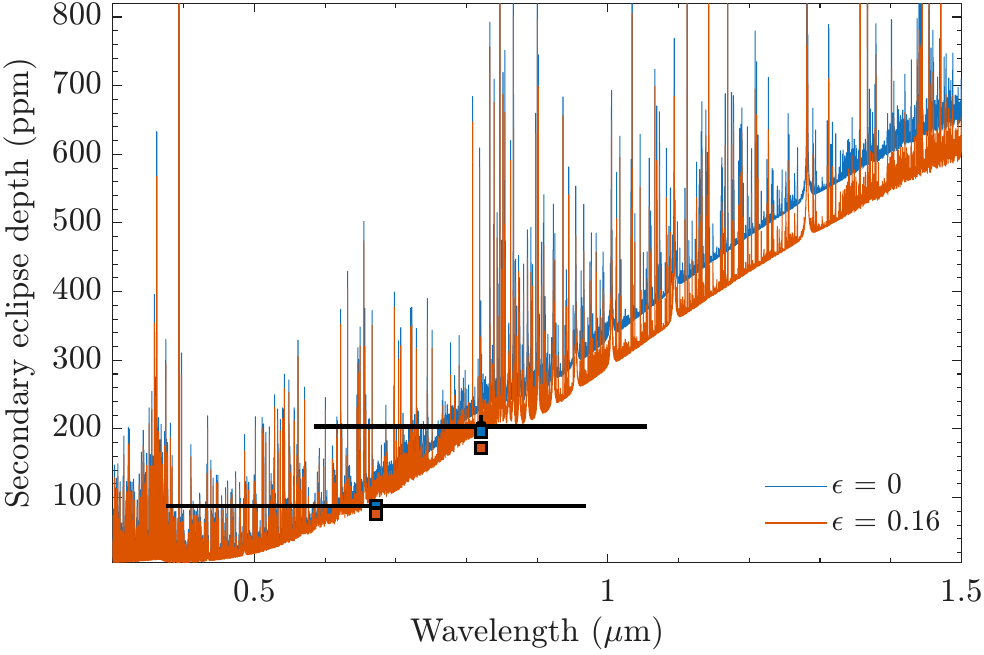}
    \caption{Theoretical occultation spectra of WASP-189\,b for two different values of the heat-redistribution factor $\epsilon$. The squares represent the bandpass-integrated occultation depths in the CHEOPS and TESS bandpasses based on the theoretical model calculations. The observational data is shown in black.} \label{fig:helios_spectra}
\end{figure} 

The resulting spectra and bandpass-integrated occultation depths for the CHEOPS and TESS filters are presented in Fig.~\ref{fig:helios_spectra}. The results indicate that very low heat redistribution efficiencies are consistent with the measured occultation depths in both bandpasses. This agrees with the findings of \citet{2020A&A...643A..94L}, who analysed only the CHEOPS occultation data.

The spectra are dominated by the free–free and bound–free continuum absorption of the H$^-$ anion, as expected for ultra-hot Jupiter atmospheres \citep[e.g.,][]{Kitzmann2018ApJ...863..183K, Lothringer2018ApJ...866...27L}. This continuum opacity is so strong that Rayleigh scattering becomes almost negligible at the wavelengths probed by CHEOPS and TESS. Consequently, the predicted geometric albedo in the CHEOPS and TESS bandpasses for the $\epsilon=0$ model is only $\sim$0.002 and $\sim10^{-4}$, respectively. Furthermore, the spectra exhibit emission features across all lines, consistent with a strong temperature inversion.

\subsubsection{Thermal structure of the atmosphere}\label{subsubsec:thermal_structure}

One of the advantages of using the phase curve model of \citet{2008ApJ...678L.129C} is that we can invert the model (Eq.~\ref{eq:cowan_agol_08}) to find the planetary brightness as a function of longitude ($\phi$) and latitude ($\theta$), $I(\phi, \theta)$. First, we can compute a longitudinal surface brightness map, $J(\phi)$, as \citep{2008ApJ...678L.129C, 2023Natur.620...67K},

\begin{equation}\label{eq:cowan_to_map}
    J(\phi) = A_0 + A_1\cos{\phi} + B_1\sin{\phi} + A_2\cos{2\phi} + B_2\sin{2\phi}.
\end{equation}

\noindent Here,

\begin{equation}
    \begin{split}
        A_0 &= (E - C_1 - C_2)/2\\ 
        A_1 &= 2C_1/\pi\\
        B_1 &= -2D_1/\pi\\
        A_2 &= 3C_2/2\\
        B_2 &= -3D_2/2.
    \end{split}
\end{equation}

\noindent Here, $E$, $C_1$, $D_1$, $C_2$ and $D_2$ are phase curve parameters from Eq.~\ref{eq:cowan_agol_08}. We note here that our phase curve observations can only estimate the longitudinal distribution of the planetary flux, while the latitudinal distribution remains unconstrained. We can make several assumptions, such as negligible brightness from the planetary poles and sinusoidal variation of brightness along latitude, to obtain a 2D brightness map, $I(\phi, \theta)$. Following \citet{2018NatAs...2..220D, 2019NatAs...3.1092K},

\begin{equation}\label{eq:2d_intensity}
    I(\phi,\theta) = \frac{3}{4} \cdot J(\phi)\cdot \sin{(\theta + \pi/2)}.
\end{equation}

\begin{figure}
    \centering
    \includegraphics[width=\linewidth]{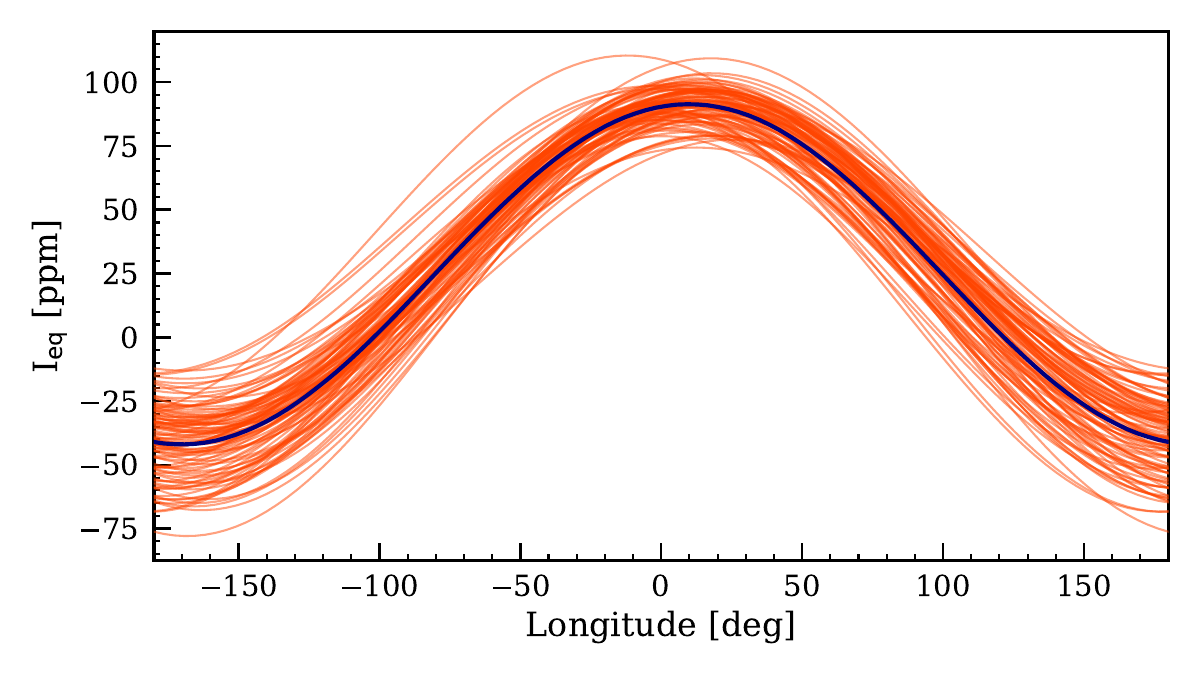}
    \includegraphics[width=\linewidth]{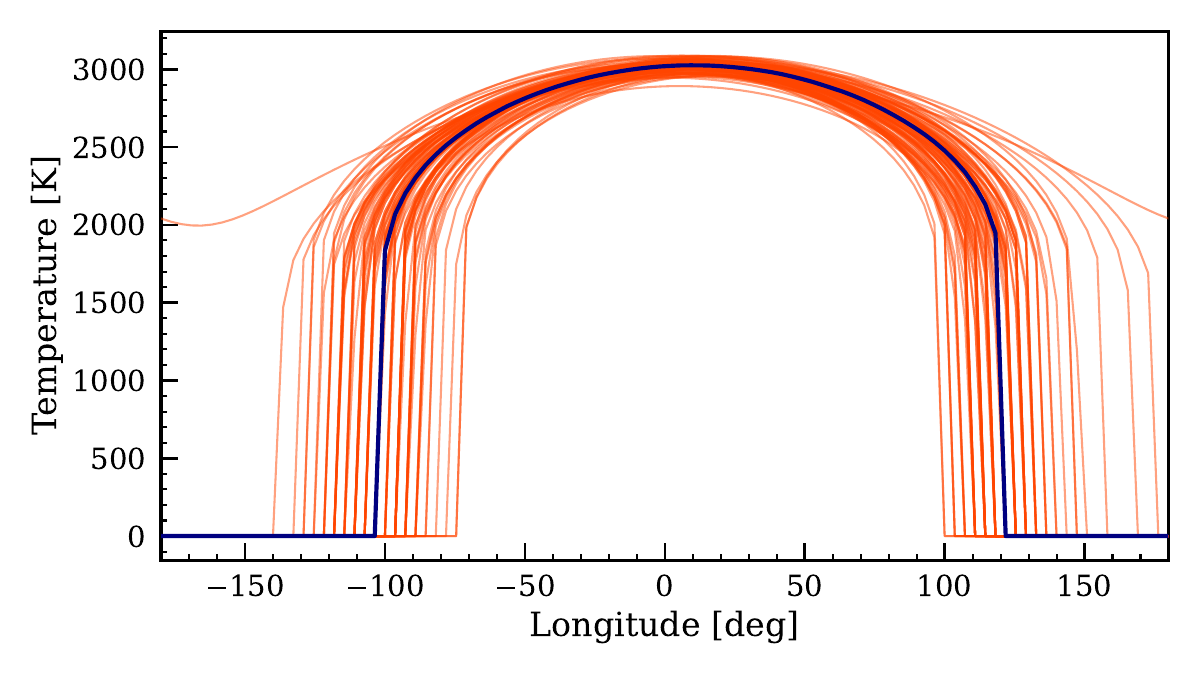}
    \includegraphics[width=\linewidth]{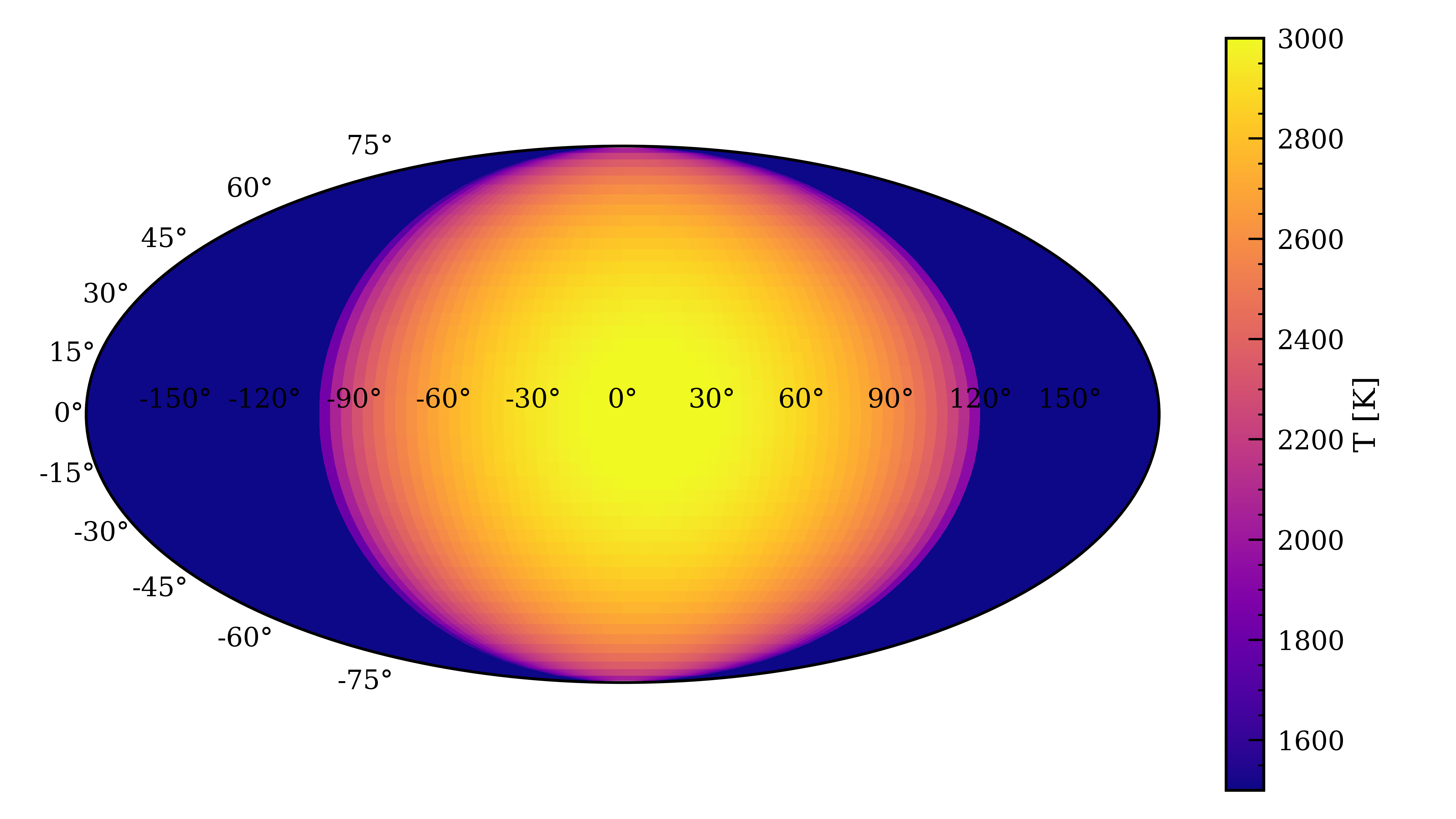}
    \caption{Brightness and temperature maps of the planet. (Top) Equatorial brightness in the units of stellar brightness as a function of longitude. (Middle) 1D temperature distribution on planetary equator as a function of planetary longitude. (Bottom) The median 2D latitude-longitude temperature map of the planet. The dark blue and orange lines in the top and middle plots show the median and randomly selected models from the posterior distribution.}
    \label{fig:temp_maps_tess}
\end{figure}

\noindent Here we add $\pi/2$ to $\theta$ in the above equation since we take $\theta=\pm\pi/2$ at poles and $\theta=0$ at the equator. The top panel of Fig.~\ref{fig:temp_maps_tess} shows the brightness variation along the equator of the planet, $I_\text{eq} = I(\phi, \theta=0)$. As can be seen, the maximum emission comes from the longitudes near the substellar point (i.e., 0$^\circ$) while the emission is consistent with zero towards nightside longitudes (i.e., near $\pm$~180$^\circ$). This shows that, as mentioned previously, there is very little phase offset and a huge contrast between the flux emitted from the dayside and the nightside.

As described in Sect.~\ref{subsubsec:atmos_model}, we mainly see planetary thermal emission in our TESS bandpass. We can use this information, along with an assumption of the planet's emission as a blackbody, to translate the derived $I(\phi, \theta)$ to a temperature map of the planet, $T_p(\phi,\theta)$. We assume that the planet emits as a blackbody without any absorption or emission due to chemical species from the atmosphere of the planet. This is a simplification, which is needed because we don't know the planetary emission spectrum. We use Planck's law to invert $I(\phi,\theta)$ to obtain $T_p(\phi,\theta)$. The $I(\phi, \theta)$ gives negative values of flux at some of the nightside longitudes, which is not physical. We set the temperature of these negative fluxes to 0~K while computing the temperature distribution. The temperature distribution at the equator is shown in the middle panel of Fig.~\ref{fig:temp_maps_tess}. As can be seen, the temperature difference between the substellar point (0$^\circ$) and the morning-evening terminator ($\pm90^\circ$) is large ($\sim1000$~K). The full median 2D temperature map derived from median $I(\phi, \theta)$ (Eq.~\ref{eq:2d_intensity}) is shown in the bottom panel of Fig.~\ref{fig:temp_maps_tess}. With certain caveats that were described above, this 2D temperature map gives a complete description of the thermal structure of the planet at the altitudes probed by TESS wavelengths. 

We can calculate the Bond albedo and heat redistribution efficiency given the 2D temperature maps. The Bond albedo can be estimated as \citep{2019NatAs...3.1092K, 2023Natur.620...67K}:

\begin{equation}\label{eq:ab}
    A_B = 1 - \frac{a^2}{\pi T_\star^4 R_\star^2} \iint T_p^4(\phi, \theta)~\sin{\theta} ~ d\theta ~ d\phi.
\end{equation}

\noindent Here, $a/R_\star$ is the scaled semi-major axis, T$_\star$ is the stellar effective temperature, and $T_p(\phi,\theta)$ is the 2D temperature distribution as a function of longitude ($\phi$) and latitude ($\theta$). We calculated A$_B$~=~\AB\, from our observations, which is mostly overestimated because of the underestimated integral in the above equation. This happens because our observations could not constrain $T_p(\phi,\theta)$ for many nightside longitudes, where we put $T_p(\phi,\theta)=0$. Therefore, the value of A$_B$ quoted above should be taken as an upper limit. We estimate a proper lower limit by recalculating A$_B$ from Eq.~\ref{eq:ab}, but this time, we replace all unconstrained $T_p(\phi,\theta)$ with the median temperature of the nightside region where we could constrain $T_p(\phi,\theta)$. We get A$_B$=~\ABNight, in this case, which is consistent with zero at 3$\sigma$. Therefore, the expected median range of A$_B$ from our observations is \ABRange, with 3$\sigma$ upper and lower limits are \ABthreeup\ and \ABthreelow, respectively.

We can estimate heat redistribution efficiency as the ratio of nightside flux to the dayside flux \citep[e.g.,][]{2022A&A...660A.123M}:

\begin{equation}\label{eq:eps}
    \varepsilon = \frac{\int_{\pi/2}^{-\pi/2} \int_{-\pi/2}^{\pi/2} F_p (\phi, \theta) \sin{(\theta + \pi/2)}~d\theta ~d\phi}{\int_{-\pi/2}^{\pi/2} \int_{-\pi/2}^{\pi/2} F_p (\phi, \theta) \sin{(\theta+\pi/2)}~d\theta ~d\phi}.
\end{equation}

\noindent Here, $F_p(\phi, \theta)$ is the total emitted flux at longitude $\phi$ and latitude $\theta$, which we can calculate as $F_p(\phi,\theta)=\sigma_{\text{sb}}T_p^4(\phi,\theta)$, where $\sigma_{\text{sb}}$ is Stefan-Boltzmann's constant. We estimate $\varepsilon$ = \epsheat\, from our observations. Again, since we cannot constrain nightside temperatures properly, this value of $\varepsilon$ is likely a lower limit. We compute an upper limit using the same method described previously in the case of A$_B$, and find $\varepsilon$ = \epsheatNight, resulting in an expected median range of \epsRange. The 3$\sigma$ upper and lower limits on $\varepsilon$ would be \epsthreeup\ and \epsthreelow, respectively. Such heat redistribution efficiency is expected for ultra-hot Jupiters, where dissociation and recombination of hydrogen increase the day-night heat transport.

We can further compute disk-integrated dayside and nightside effective temperatures of the planet \citep{2019NatAs...3.1092K},

\begin{equation}\label{eq:tday_tnight}
    \begin{split}
        T_\text{day}^4 &= \frac{1}{2\pi} \int_{-\pi/2} ^{\pi/2} \int_{-\pi/2} ^{\pi/2} T^4(\phi,\theta) \, \sin{\theta}~d\theta~d\phi \\
        T_\text{night}^4 &= \frac{1}{2\pi} \int_{\pi/2} ^{-\pi/2} \int_{-\pi/2} ^{\pi/2} T^4(\phi,\theta) \, \sin{\theta}~d\theta~d\phi
    \end{split}
\end{equation}

\noindent Here $T(\phi, \theta)$ is the 2D temperature map inverted from our observationally derived $I(\phi, \theta)$ (Eq.~\ref{eq:2d_intensity}). We estimate {\small $T_\text{day}$} = \Tday~{\small K} and {\small $T_\text{night}$} = \Tnight~{\small K}. The latter of which is likely underestimated since we cannot properly constrain the nightside flux. \citet{2019NatAs...3.1092K} studied the thermal structure of several hot Jupiters and showed that most of them show a uniform nightside temperature of about 1100~K, primarily because of the presence of clouds. With increasing stellar irradiation, the clouds can disperse, increasing the nightside temperature of highly irradiated planets. The nightside temperature of WASP-189\,b nicely fits in the trend found by \citet{2019NatAs...3.1092K}.

We finally note that the physical quantities derived here depend on the temperature map of the planet, which, in turn, depends on the phase curve model we used.

\subsubsection{On the contribution of reflected light}\label{subsubsec:cheops_pc_analysis}

\begin{figure}
    \centering
    \includegraphics[width=\linewidth]{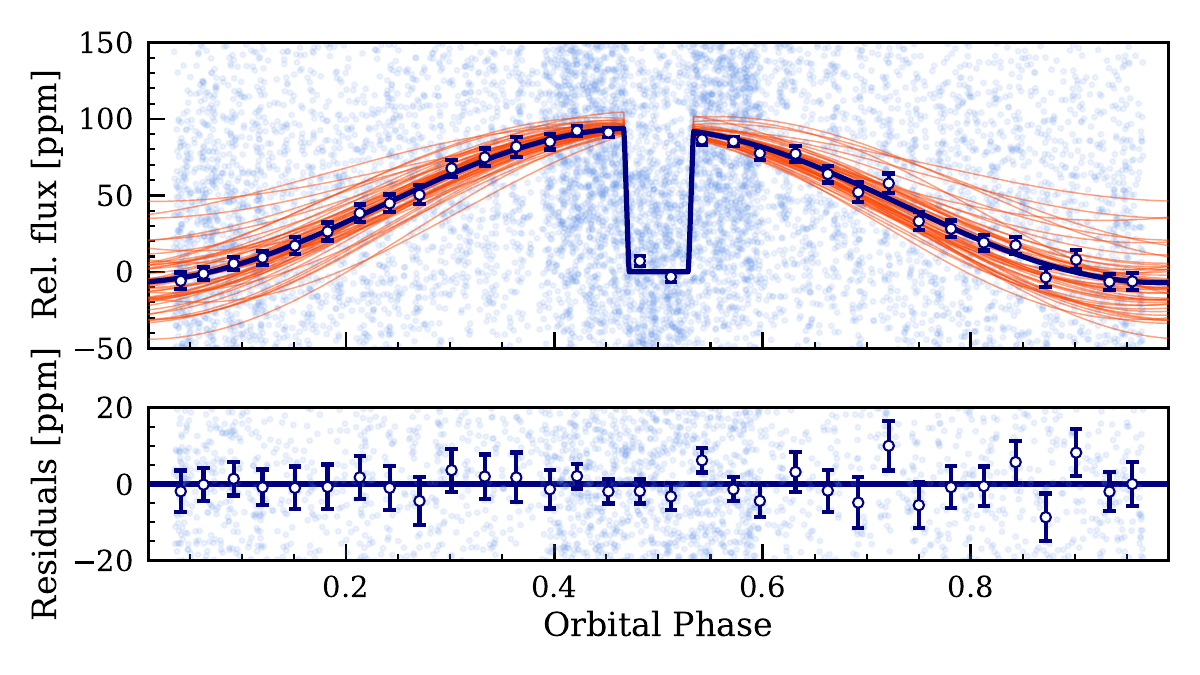}
    \includegraphics[width=\linewidth]{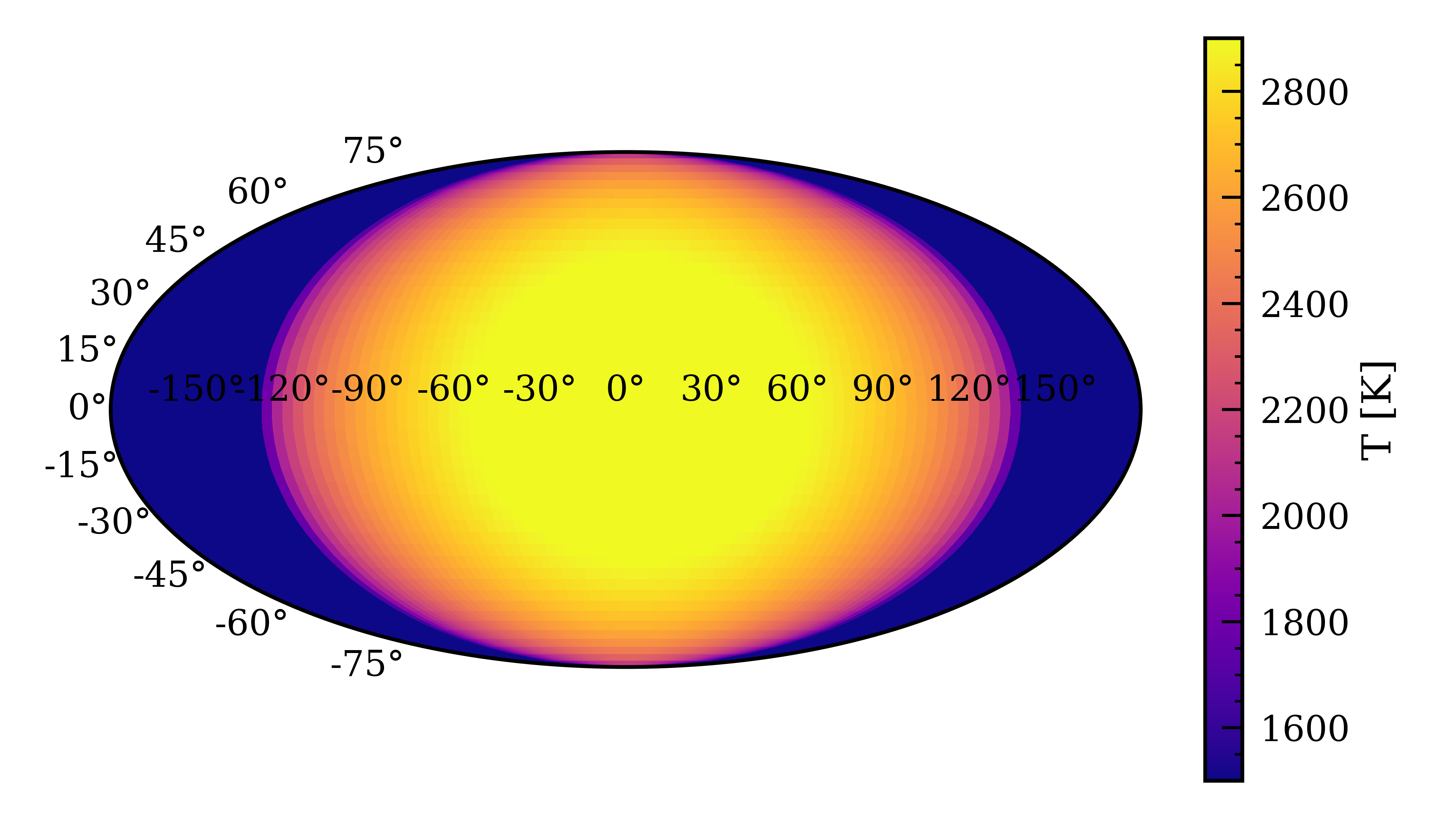}
    \caption{CHEOPS phase curve and inferred temperature map from it. (Top) The detrended and phase-folded CHEOPS data along with the median (dark blue line, top panel) and models computed from randomly selected posteriors (orange line, top panel). The bottom panel shows the residuals after subtracting the median model. The light and dark blue points show the unbinned and binned data, respectively. (Bottom) The derived median temperature map as a function of latitude and longitude.}
    \label{fig:cheops_observations}
\end{figure}

Our atmospheric modelling (Sect.~\ref{subsubsec:atmos_model}) suggests the contribution of reflected light in both TESS and CHEOPS bandpasses to be small. This is corroborated by our computation of Bond albedo, the lower limit of which we found to be consistent with zero at 3$\sigma$ (Sect.~\ref{subsubsec:thermal_structure}). The optical bandpass of CHEOPS is better suited to constrain the geometric albedo of WASP-189\,b. \citet{2022A&A...659A..74D}, who observed optical CHEOPS phase curves and occultations, suggested that their observations can be explained by $A_B=0$, in the extreme case. However, they also contemplate that the reflected light contribution to the emission in the CHEOPS bandpass could be non-zero because it would require a negative Bond albedo to obtain the observed occultation depth only from thermal emission. We argue that the negative Bond albedo arises because of the inherently flawed nature of their 0D model \citep[from][]{2011ApJ...729...54C}. \citet{2022A&A...660A.123M} showed that such 0D models can give negative Bond albedos in certain situations because of global non-conservation of energy. They further recommended using 2D temperature maps to calculate Bond albedo. Indeed, we show that our phase curve model, and subsequently derived temperature maps, when applied to CHEOPS observations, give a small non-negative Bond albedo as expected.

For the purpose of this analysis, we analysed four occultations and two phase-curve observations from CHEOPS published by \citet{2020A&A...643A..94L, 2022A&A...659A..74D}. \boldchange{This analysis was performed using the CHEOPS data alone.} Since we are only interested in phase curves, we masked all transits present in the data. Our planetary model is the same as the one described in Sect.~\ref{subsec:pc_model}. There is a strong instrumental noise in CHEOPS data as mentioned by \citet{2022A&A...659A..74D}. We fit a model similar to \citet{2022A&A...659A..74D} to model the instrumental noise. This model included a Fourier series up to fifth order for occultations and phase curves to fit roll-angle modulation, linear (for phase curves) and quadratic polynomials (for occultations) in time to model long-term trends, and a linear model in \texttt{thermFront\_2} parameter to account for the flux ramp. Finally, we fitted a GP model built from an SHO kernel implemented in \texttt{celerite2} \citep{celerite1, celerite2} to phase curve data to model the stellar noise \citep{2022A&A...659A..74D}. We fixed all planetary parameters, except for phase curve parameters, to their values from \citet{2022A&A...659A..74D} in this analysis.

The detrended data along with the median and the models generated from the randomly selected posteriors are shown in the top panel of Fig.~\ref{fig:cheops_observations}. Our constrained values of occultation depth, nightside flux and phase offset, {\small $93.9^{+2.1}_{-1.9}$~ppm}, {\small $-6.7^{+15.9}_{-17.1}$~ppm}, and {\small $-5.6^{+8.7}_{-6.9}$~deg}, are consistent with their values from \citet{2022A&A...659A..74D}. We get smaller uncertainties on these parameters because we fixed planetary parameters in our analysis. We then used the formalism presented in Sect.~\ref{subsubsec:thermal_structure} to invert this phase curve to find the brightness distribution, $I(\phi, \theta)$, and temperature distribution of the planet. The median temperature distribution is shown in the bottom panel of Fig.~\ref{fig:cheops_observations}. Again, where the planetary flux is not properly constrained, on the nightside where we get $I(\phi, \theta)<0$ for some longitudes, we set $T_p(\phi,\theta)=0$ (e.g., dark blue bands in Fig.~\ref{fig:cheops_observations}). We estimate the Bond albedo, $A_B = {\small 0.19 ^{+0.06}_{-0.10}}$, from this temperature map, which we found to be consistent with zero at $\sim2\sigma$. As we cannot constrain the nightside temperatures properly, the real value of $A_B$ is likely to be much lower than the value we report, which is imprinted in its large lower uncertainty. To properly account for this, we set the temperature of the unconstrained nightside region to the median nightside temperature. We get $A_B = {\small 0.05 ^{+0.04}_{-0.03}}$, making the expected median range of A$_B$, {\small 0.05-0.19}. The 3$\sigma$ upper and lower limits of $A_B$ are 0.37 and $-$0.04, respectively. Such low values of Bond albedo could mean that the contribution of reflective light in the CHEOPS bandpass is minimal. This is also hinted at by our atmospheric modelling (see Sect.~\ref{subsubsec:atmos_model}). However, it is hard to calculate the exact contribution of reflected light in both TESS and CHEOPS bandpasses in the absence of a true spectrum.

\subsubsection{Bandpass dependent thermal structure}\label{subsubsec:tess_cheops_comp}

\begin{figure}
    \centering
    \includegraphics[width=\linewidth]{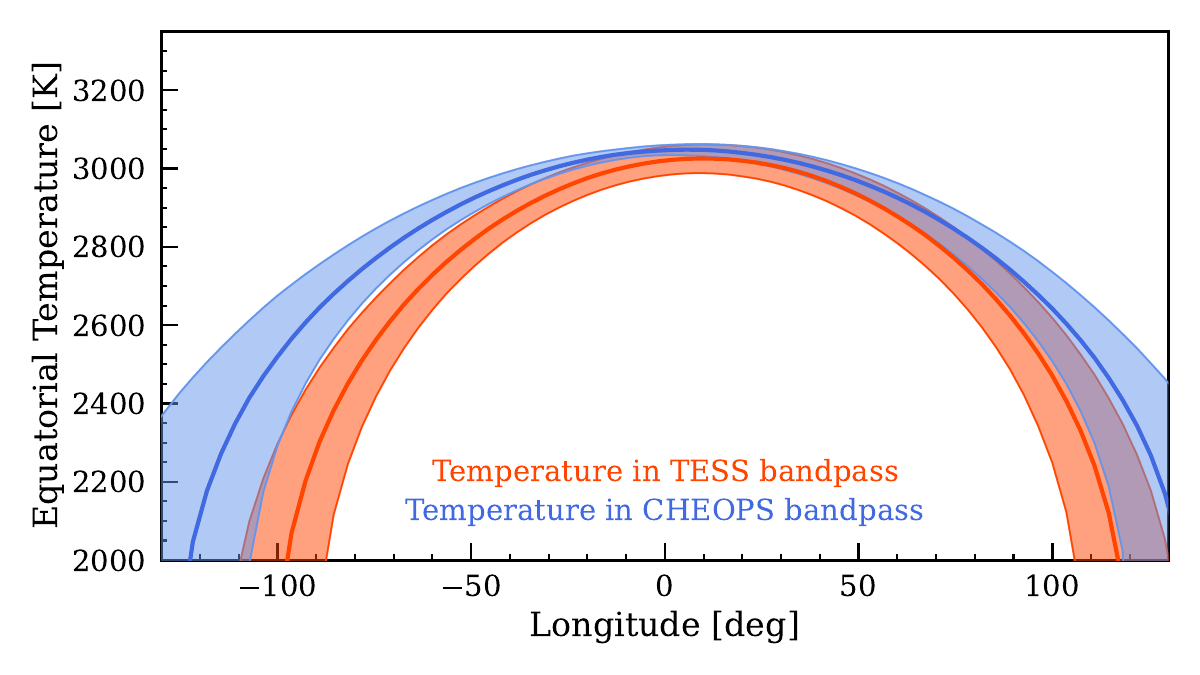}
    \caption{Equatorial temperatures as a function of longitude as observed by TESS (in orange) and CHEOPS (in blue) bandpasses. The solid lines give median values of temperatures, while the shaded regions show the bands of 1$\sigma$ uncertainty in temperatures. We do not plot some nightside longitudes since our modelling cannot properly constrain temperatures on very high longitudes.}
    \label{fig:temp_map_comparison}
\end{figure}

We computed the temperature distribution of WASP-189\,b in two different bandpasses. TESS and CHEOPS bandpasses, while overlapping for large wavelengths, are sensitive to slightly different wavelengths, with the CHEOPS bandpass being more sensitive to bluer wavelengths than the TESS bandpass. They naturally probe different atmospheric layers in the atmosphere of the planet. Comparing temperature distribution constrained by both bandpasses can, thus, give us how the thermal structure varies with altitude in the atmosphere.

Fig.~\ref{fig:temp_map_comparison} shows the equatorial temperatures as a function of longitude as observed by TESS and CHEOPS. As can be seen, the estimated temperatures are almost similar in both bandpasses. However, the temperature is slightly higher in the CHEOPS bandpass compared to the TESS bandpass. This could be explained by the presence of short-wave absorbers such as titanium oxide (TiO) in the atmosphere of WASP-189\,b \citep{2022NatAs...6..449P, 2023A&A...678A.182P}. Such short-wave absorbers can create a temperature inversion in the atmosphere, which was indeed empirically found by \citet{2025A&A...693A..72L} for WASP-189\,b. The existence of short-wave absorbers also means that the radiation detected by the bluer bandpass of CHEOPS is radiated by the upper, hotter layers. On the other hand, TESS's redder bandpass receives radiation from slightly deeper, and thus, cooler, layers. This could explain the temperature difference in the CHEOPS and TESS bandpasses. Given that both bandpasses overlap for a large wavelength range, the temperature difference is small, as seen in Fig.~\ref{fig:temp_map_comparison}.

\section{Stellar variability}\label{sec:stel_var}

\begin{figure}
    \centering
    \includegraphics[width=\linewidth]{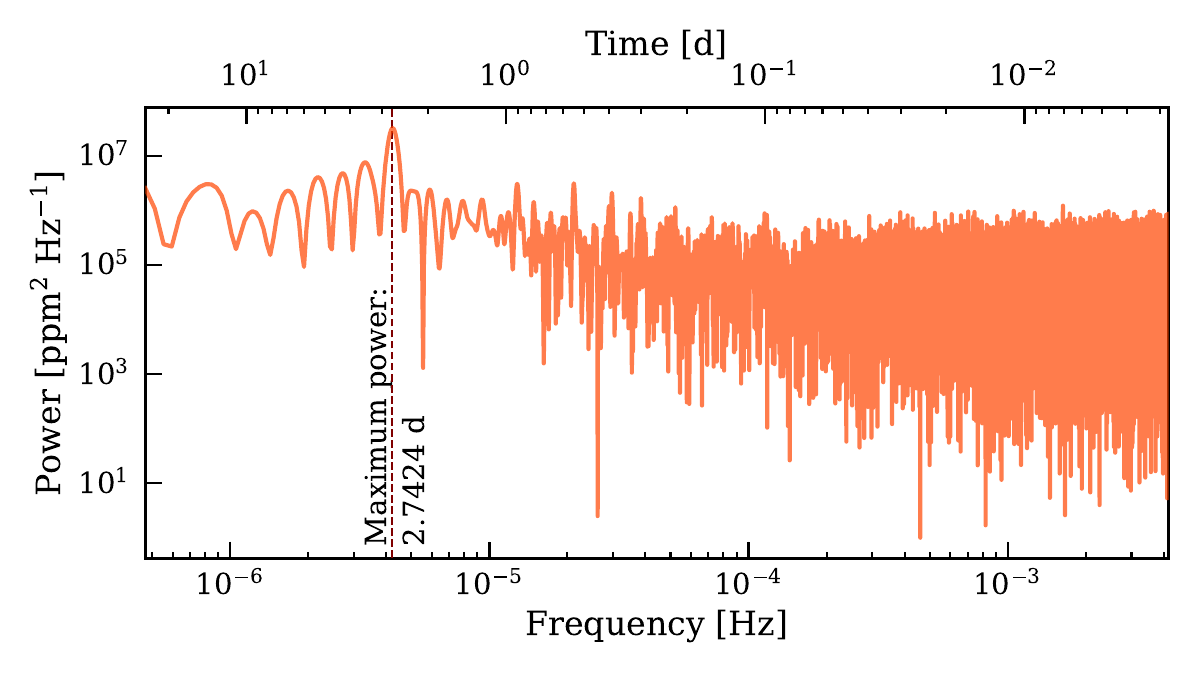}
    \caption{Lomb-Scargle periodogram of the TESS light curve after subtracting the instrumental noise. The bottom and upper axes of the figure show the frequency and time domain, respectively. The period range covered by the periodogram is from the period corresponding to the Nyquist frequency to the total duration of the observation.}
    \label{fig:psd_tess}
\end{figure}

\citet{2022A&A...659A..74D} found photometric variability in the CHEOPS phase curve of WASP-189\,b with the period of about 1.3~days. They attributed this photometric variability to stellar rotation since the period of variability was very close to the stellar rotation period, $P_\star$ = \Pstar~{\small d}. We recovered this photometric variability in our CHEOPS phase curve analysis presented in Sect.~\ref{subsubsec:cheops_pc_analysis} (see the bottom panel of Fig.~\ref{fig:GP_models} for our best-fitted GP model to CHEOPS data). We fit a slightly larger undamped period of the process compared to \citet{2022A&A...659A..74D} (see Table~\ref{tab:GP_fitted_params}). Given the presence of photometric variability in CHEOPS data, we searched for the same in the TESS data. The longer observing baseline the TESS data provides is especially suitable for determining the long-term variability of WASP-189. Fig.~\ref{fig:psd_tess} shows the Lomb-Scargle periodogram of the TESS data after subtracting the instrumental noise and masking transits and occultations. The periodogram has one significant peak at \boldchange{\psdpeak~d}, which, although much larger than the period of photometric variability found by \citet{2022A&A...659A..74D}, is consistent with the orbital period of the planet (P = 2.72~d). We, therefore, think that the peak corresponds to the planetary phase curve signal rather than any stellar variability. Curiously, unlike \citet{2022A&A...659A..74D}, there are no significant peaks around 1.2~days. The photometric variability discovered by CHEOPS seems to be absent in the TESS data. We note here that the TESS data is less precise compared to the CHEOPS data: we estimate the photometric precision of both datasets by calculating MAD of residuals, found by subtracting best-fitted model from the raw flux, binned over 1~hr. We found \phometricprecCHEOPS\ and \phometricprecTESS\ in 1~hr for CHEOPS and TESS datasets, respectively. It is possible that the stellar photometric variability is not visible because of noise in the TESS data. 

Although we did not detect short-term variability akin to that found in CHEOPS, the periodogram shows enhanced power at lower frequencies, indicating the presence of long-term noise. Given the amount of noise in the raw data, this is likely leftover instrumental noise. Therefore, we added a GP model built from an SHO kernel to model the additional correlated noise (see Sect.~\ref{subsec:misc_post} for the details). The best-fitted GP model is shown in the top panel of Fig.~\ref{fig:GP_models}. The GP model can produce oscillations if the quality factor, Q, of the GP is greater than 0.5 (i.e, $\ln{Q}>-0.69$). The fitted Q value of the GP is below this threshold (see Table~\ref{tab:GP_fitted_params}), so we do not see regular oscillations in the model. The best-fitted undamped period of the GP is \GPundamped~{\small d}, which gives the damped period of \GPdamped~{\small d} (or, \GPdampedhr~{\small hr}). Although the stellar properties of WASP-189 put it in a parameter space where $\delta$-Scuti and $\gamma$-Doradus stars are found, the irregularities of the fitted GP model mean that the noise is not likely of stellar origin.

\section{Conclusions}\label{sec:conclusions}

We analysed the new TESS and archival CHEOPS phase curves to investigate the orbital architecture and thermal structure of an ultra-hot Jupiter WASP-189\,b.

\begin{enumerate}
    \item Our analysis of gravity-darkened transit allowed us to examine the orbital architecture. We found that, in agreement with \citet{2022A&A...659A..74D}, the planet is in a polar orbit around its host star, with the 3D spin-orbit angle being \PSI~{\small deg}.

    \item The whole dataset, including archival CHEOPS data and TESS data, spans over roughly two years. We used this data over a long baseline to see if there is orbital precession because of stellar oblateness. We found that our impact parameter and projected orbital obliquity are consistent with their corresponding values from archival CHEOPS data, with their rate of changes, \bbdot~yr$^{-1}$ and \lampdot~deg~yr$^{-1}$, respectively, consistent with zero. This suggests an absence of a fast orbital precession.

    \item We detected a significant phase curve with an occultation depth of \ECowan~{\small ppm}, a negative, yet consistent with zero, nightside flux (F$_n$/F$_\star$ = \FnFs~{\small ppm}), and a negligible phase offset of \phioff~{\small deg}. Our atmospheric forward models of thermal emission with minor contribution from reflected light in an atmosphere with inefficient heat transport can explain the observed occultation depths in both TESS and CHEOPS bandpasses.

    \item Our phase-curve model enabled us to invert the phase curve to find temperature maps of the planet, which we calculated for both TESS and CHEOPS bandpasses. Upon comparing temperature maps in both bandpasses, we found that, while the temperatures are almost similar in both bands, it is slightly higher in the CHEOPS band compared to the TESS band throughout the dayside. A temperature inversion, caused by previously detected short-wave absorbers such as TiO, could readily explain this \citep{2022NatAs...6..449P, 2023A&A...678A.182P}.

    \item We used temperature maps to estimate the Bond albedo ($A_B$) in TESS and CHEOPS bandpasses. Given the unconstrained nightside temperatures, we could derive accepted median ranges of $A_B$, \ABRange\, in the TESS bandpass and {\small 0.05-0.19} in the CHEOPS bandpass. We also derived 3$\sigma$ upper and lower limits of $A_B$ for both bandpasses. Similarly, we constrained the heat re-distribution efficiency in the \epsRange\ median range for the TESS bandpass.

    \item \citet{2022A&A...659A..74D} detected a long-term photometric variability at the period of about 1.3~days, which they attributed to the stellar rotation. We did not detect a similar photometric variability in our TESS observations. This may be because of the lower photometric precision of the TESS data compared to the CHEOPS observations. However, we detected correlated noise of unknown origin. 

\end{enumerate}

Our phase curve observations can primarily constrain the longitudinal thermal structure, and we made additional assumptions to obtain a 2D latitude-longitude thermal structure of the atmosphere. However, it is possible to directly constrain 2D latitude-longitude thermal structure with ultra-precise occultation observations. Moreover, spectroscopic observations can provide altitude information, which helps to understand 3D atmospheric structure. Observations with the James Webb Space Telescope can achieve this \citep[][]{2024AJ....168....4H}. Such a 3D atmospheric structure is crucial in studying not only the thermal structure but also atmospheric dynamics.

\begin{acknowledgements}
We would like to thank an anonymous referee for their detailed referee report and suggestions which significantly improved the manuscript.
JAP and ABr were supported by SNSA.
DK acknowledges financial support from the Center for Space and Habitability (CSH) of the University of Bern.
TWi acknowledges support from the UKSA and the University of Warwick.
ADe has received funding from the Swiss National Science Foundation for project 200021\_200726
ML acknowledges support of the Swiss National Science Foundation under grant number PCEFP2\_194576. The contributions of ML and ADe have been carried out within the framework of the NCCR PlanetS supported by the Swiss National Science Foundation under grant 51NF40\_205606.
VS acknowledge support from CHEOPS ASI-INAF agreement no. 2019-29-HH.0.
\end{acknowledgements}

\bibliographystyle{aa} % style aa.bst
\bibliography{references} % your references Yourfile.bib

@ARTICLE{2021ExA....51..109B,
       author = {{Benz}, W. and {Broeg}, C. and {Fortier}, A. and {Rando}, N. and {Beck}, T. and {Beck}, M. and {Queloz}, D. and {Ehrenreich}, D. and {Maxted}, P.~F.~L. and {Isaak}, K.~G. and {Billot}, N. and {Alibert}, Y. and {Alonso}, R. and {Ant{\'o}nio}, C. and {Asquier}, J. and {Bandy}, T. and {B{\'a}rczy}, T. and {Barrado}, D. and {Barros}, S.~C.~C. and {Baumjohann}, W. and {Bekkelien}, A. and {Bergomi}, M. and {Biondi}, F. and {Bonfils}, X. and {Borsato}, L. and {Brandeker}, A. and {Busch}, M. -D. and {Cabrera}, J. and {Cessa}, V. and {Charnoz}, S. and {Chazelas}, B. and {Collier Cameron}, A. and {Corral Van Damme}, C. and {Cortes}, D. and {Davies}, M.~B. and {Deleuil}, M. and {Deline}, A. and {Delrez}, L. and {Demangeon}, O. and {Demory}, B.~O. and {Erikson}, A. and {Farinato}, J. and {Fossati}, L. and {Fridlund}, M. and {Futyan}, D. and {Gandolfi}, D. and {Garcia Munoz}, A. and {Gillon}, M. and {Guterman}, P. and {Gutierrez}, A. and {Hasiba}, J. and {Heng}, K. and {Hernandez}, E. and {Hoyer}, S. and {Kiss}, L.~L. and {Kovacs}, Z. and {Kuntzer}, T. and {Laskar}, J. and {Lecavelier des Etangs}, A. and {Lendl}, M. and {L{\'o}pez}, A. and {Lora}, I. and {Lovis}, C. and {L{\"u}ftinger}, T. and {Magrin}, D. and {Malvasio}, L. and {Marafatto}, L. and {Michaelis}, H. and {de Miguel}, D. and {Modrego}, D. and {Munari}, M. and {Nascimbeni}, V. and {Olofsson}, G. and {Ottacher}, H. and {Ottensamer}, R. and {Pagano}, I. and {Palacios}, R. and {Pall{\'e}}, E. and {Peter}, G. and {Piazza}, D. and {Piotto}, G. and {Pizarro}, A. and {Pollaco}, D. and {Ragazzoni}, R. and {Ratti}, F. and {Rauer}, H. and {Ribas}, I. and {Rieder}, M. and {Rohlfs}, R. and {Safa}, F. and {Salatti}, M. and {Santos}, N.~C. and {Scandariato}, G. and {S{\'e}gransan}, D. and {Simon}, A.~E. and {Smith}, A.~M.~S. and {Sordet}, M. and {Sousa}, S.~G. and {Steller}, M. and {Szab{\'o}}, G.~M. and {Szoke}, J. and {Thomas}, N. and {Tschentscher}, M. and {Udry}, S. and {Van Grootel}, V. and {Viotto}, V. and {Walter}, I. and {Walton}, N.~A. and {Wildi}, F. and {Wolter}, D.},
        title = "{The CHEOPS mission}",
      journal = {Experimental Astronomy},
         year = 2021,
        month = feb,
       volume = {51},
       number = {1},
        pages = {109-151},
          doi = {10.1007/s10686-020-09679-4},
archivePrefix = {arXiv},
       eprint = {2009.11633},
 primaryClass = {astro-ph.IM},
       adsurl = {https://ui.adsabs.harvard.edu/abs/2021ExA....51..109B}
}

@INPROCEEDINGS{2016SPIE.9913E..3EJ,
       author = {{Jenkins}, Jon M. and {Twicken}, Joseph D. and {McCauliff}, Sean and {Campbell}, Jennifer and {Sanderfer}, Dwight and {Lung}, David and {Mansouri-Samani}, Masoud and {Girouard}, Forrest and {Tenenbaum}, Peter and {Klaus}, Todd and {Smith}, Jeffrey C. and {Caldwell}, Douglas A. and {Chacon}, A.~D. and {Henze}, Christopher and {Heiges}, Cory and {Latham}, David W. and {Morgan}, Edward and {Swade}, Daryl and {Rinehart}, Stephen and {Vanderspek}, Roland},
        title = "{The TESS science processing operations center}",
    booktitle = {Software and Cyberinfrastructure for Astronomy IV},
         year = 2016,
       editor = {{Chiozzi}, Gianluca and {Guzman}, Juan C.},
       series = {Society of Photo-Optical Instrumentation Engineers (SPIE) Conference Series},
       volume = {9913},
        month = aug,
          eid = {99133E},
        pages = {99133E},
          doi = {10.1117/12.2233418},
       adsurl = {https://ui.adsabs.harvard.edu/abs/2016SPIE.9913E..3EJ}
}

@ARTICLE{2012PASP..124.1000S,
       author = {{Smith}, Jeffrey C. and {Stumpe}, Martin C. and {Van Cleve}, Jeffrey E. and {Jenkins}, Jon M. and {Barclay}, Thomas S. and {Fanelli}, Michael N. and {Girouard}, Forrest R. and {Kolodziejczak}, Jeffery J. and {McCauliff}, Sean D. and {Morris}, Robert L. and {Twicken}, Joseph D.},
        title = "{Kepler Presearch Data Conditioning II - A Bayesian Approach to Systematic Error Correction}",
      journal = {\pasp},
         year = 2012,
        month = sep,
       volume = {124},
       number = {919},
        pages = {1000},
          doi = {10.1086/667697},
archivePrefix = {arXiv},
       eprint = {1203.1383},
 primaryClass = {astro-ph.IM},
       adsurl = {https://ui.adsabs.harvard.edu/abs/2012PASP..124.1000S}
}

@ARTICLE{2014PASP..126..100S,
       author = {{Stumpe}, Martin C. and {Smith}, Jeffrey C. and {Catanzarite}, Joseph H. and {Van Cleve}, Jeffrey E. and {Jenkins}, Jon M. and {Twicken}, Joseph D. and {Girouard}, Forrest R.},
        title = "{Multiscale Systematic Error Correction via Wavelet-Based Bandsplitting in Kepler Data}",
      journal = {\pasp},
         year = 2014,
        month = jan,
       volume = {126},
       number = {935},
        pages = {100},
          doi = {10.1086/674989},
       adsurl = {https://ui.adsabs.harvard.edu/abs/2014PASP..126..100S}
}

@ARTICLE{2022A&A...659A..74D,
       author = {{Deline}, A. and {Hooton}, M.~J. and {Lendl}, M. and {Morris}, B. and {Salmon}, S. and {Olofsson}, G. and {Broeg}, C. and {Ehrenreich}, D. and {Beck}, M. and {Brandeker}, A. and {Hoyer}, S. and {Sulis}, S. and {Van Grootel}, V. and {Bourrier}, V. and {Demangeon}, O. and {Demory}, B. -O. and {Heng}, K. and {Parviainen}, H. and {Serrano}, L.~M. and {Singh}, V. and {Bonfanti}, A. and {Fossati}, L. and {Kitzmann}, D. and {Sousa}, S.~G. and {Wilson}, T.~G. and {Alibert}, Y. and {Alonso}, R. and {Anglada}, G. and {B{\'a}rczy}, T. and {Barrado Navascues}, D. and {Barros}, S.~C.~C. and {Baumjohann}, W. and {Beck}, T. and {Bekkelien}, A. and {Benz}, W. and {Billot}, N. and {Bonfils}, X. and {Cabrera}, J. and {Charnoz}, S. and {Collier Cameron}, A. and {Corral van Damme}, C. and {Csizmadia}, Sz. and {Davies}, M.~B. and {Deleuil}, M. and {Delrez}, L. and {de Roche}, T. and {Erikson}, A. and {Fortier}, A. and {Fridlund}, M. and {Futyan}, D. and {Gandolfi}, D. and {Gillon}, M. and {G{\"u}del}, M. and {Gutermann}, P. and {Hasiba}, J. and {Isaak}, K.~G. and {Kiss}, L. and {Laskar}, J. and {Lecavelier des Etangs}, A. and {Lovis}, C. and {Magrin}, D. and {Maxted}, P.~F.~L. and {Munari}, M. and {Nascimbeni}, V. and {Ottensamer}, R. and {Pagano}, I. and {Pall{\'e}}, E. and {Peter}, G. and {Piotto}, G. and {Pollacco}, D. and {Queloz}, D. and {Ragazzoni}, R. and {Rando}, N. and {Rauer}, H. and {Ribas}, I. and {Santos}, N.~C. and {Scandariato}, G. and {S{\'e}gransan}, D. and {Simon}, A.~E. and {Smith}, A.~M.~S. and {Steller}, M. and {Szab{\'o}}, Gy. M. and {Thomas}, N. and {Udry}, S. and {Walter}, I. and {Walton}, N.},
        title = "{The atmosphere and architecture of WASP-189 b probed by its CHEOPS phase curve}",
      journal = {\aap},
         year = 2022,
        month = mar,
       volume = {659},
          eid = {A74},
        pages = {A74},
          doi = {10.1051/0004-6361/202142400},
archivePrefix = {arXiv},
       eprint = {2201.04518},
 primaryClass = {astro-ph.EP},
       adsurl = {https://ui.adsabs.harvard.edu/abs/2022A&A...659A..74D}
}

@ARTICLE{2020A&A...643A..94L,
       author = {{Lendl}, M. and {Csizmadia}, Sz. and {Deline}, A. and {Fossati}, L. and {Kitzmann}, D. and {Heng}, K. and {Hoyer}, S. and {Salmon}, S. and {Benz}, W. and {Broeg}, C. and {Ehrenreich}, D. and {Fortier}, A. and {Queloz}, D. and {Bonfanti}, A. and {Brandeker}, A. and {Collier Cameron}, A. and {Delrez}, L. and {Garcia Mu{\~n}oz}, A. and {Hooton}, M.~J. and {Maxted}, P.~F.~L. and {Morris}, B.~M. and {Van Grootel}, V. and {Wilson}, T.~G. and {Alibert}, Y. and {Alonso}, R. and {Asquier}, J. and {Bandy}, T. and {B{\'a}rczy}, T. and {Barrado}, D. and {Barros}, S.~C.~C. and {Baumjohann}, W. and {Beck}, M. and {Beck}, T. and {Bekkelien}, A. and {Bergomi}, M. and {Billot}, N. and {Biondi}, F. and {Bonfils}, X. and {Bourrier}, V. and {Busch}, M. -D. and {Cabrera}, J. and {Cessa}, V. and {Charnoz}, S. and {Chazelas}, B. and {Corral Van Damme}, C. and {Davies}, M.~B. and {Deleuil}, M. and {Demangeon}, O.~D.~S. and {Demory}, B. -O. and {Erikson}, A. and {Farinato}, J. and {Fridlund}, M. and {Futyan}, D. and {Gandolfi}, D. and {Gillon}, M. and {Guterman}, P. and {Hasiba}, J. and {Hernandez}, E. and {Isaak}, K.~G. and {Kiss}, L. and {Kuntzer}, T. and {Lecavelier des Etangs}, A. and {L{\"u}ftinger}, T. and {Laskar}, J. and {Lovis}, C. and {Magrin}, D. and {Malvasio}, L. and {Marafatto}, L. and {Michaelis}, H. and {Munari}, M. and {Nascimbeni}, V. and {Olofsson}, G. and {Ottacher}, H. and {Ottensamer}, R. and {Pagano}, I. and {Pall{\'e}}, E. and {Peter}, G. and {Piazza}, D. and {Piotto}, G. and {Pollacco}, D. and {Ratti}, F. and {Rauer}, H. and {Ragazzoni}, R. and {Rando}, N. and {Ribas}, I. and {Rieder}, M. and {Rohlfs}, R. and {Safa}, F. and {Santos}, N.~C. and {Scandariato}, G. and {S{\'e}gransan}, D. and {Simon}, A.~E. and {Singh}, V. and {Smith}, A.~M.~S. and {Sordet}, M. and {Sousa}, S.~G. and {Steller}, M. and {Szab{\'o}}, Gy. M. and {Thomas}, N. and {Tschentscher}, M. and {Udry}, S. and {Viotto}, V. and {Walter}, I. and {Walton}, N.~A. and {Wildi}, F. and {Wolter}, D.},
        title = "{The hot dayside and asymmetric transit of WASP-189 b seen by CHEOPS}",
      journal = {\aap},
         year = 2020,
        month = nov,
       volume = {643},
          eid = {A94},
        pages = {A94},
          doi = {10.1051/0004-6361/202038677},
archivePrefix = {arXiv},
       eprint = {2009.13403},
 primaryClass = {astro-ph.EP},
       adsurl = {https://ui.adsabs.harvard.edu/abs/2020A&A...643A..94L}
}

@ARTICLE{2018arXiv180904897A,
       author = {{Anderson}, D.~R. and {Temple}, L.~Y. and {Nielsen}, L.~D. and {Burdanov}, A. and {Hellier}, C. and {Bouchy}, F. and {Brown}, D.~J.~A. and {Collier Cameron}, A. and {Gillon}, M. and {Jehin}, E. and {Maxted}, P.~F.~L. and {Pepe}, F. and {Pollacco}, D. and {Pozuelos}, F.~J. and {Queloz}, D. and {S{\'e}gransan}, D. and {Smalley}, B. and {Triaud}, A.~H.~M.~J. and {Turner}, O.~D. and {Udry}, S. and {West}, R.~G.},
        title = "{WASP-189b: an ultra-hot Jupiter transiting the bright A star HR 5599 in a polar orbit}",
      journal = {arXiv e-prints},
         year = 2018,
        month = sep,
          eid = {arXiv:1809.04897},
        pages = {arXiv:1809.04897},
          doi = {10.48550/arXiv.1809.04897},
archivePrefix = {arXiv},
       eprint = {1809.04897},
 primaryClass = {astro-ph.EP},
       adsurl = {https://ui.adsabs.harvard.edu/abs/2018arXiv180904897A}
}

@ARTICLE{1924MNRAS..84..684V,
       author = {{von Zeipel}, H.},
        title = "{The radiative equilibrium of a slightly oblate rotating star}",
      journal = {\mnras},
         year = 1924,
        month = jun,
       volume = {84},
        pages = {684-701},
          doi = {10.1093/mnras/84.9.684},
       adsurl = {https://ui.adsabs.harvard.edu/abs/1924MNRAS..84..684V}
}

@ARTICLE{2009ApJ...705..683B,
       author = {{Barnes}, Jason W.},
        title = "{Transit Lightcurves of Extrasolar Planets Orbiting Rapidly Rotating Stars}",
      journal = {\apj},
         year = 2009,
        month = nov,
       volume = {705},
       number = {1},
        pages = {683-692},
          doi = {10.1088/0004-637X/705/1/683},
archivePrefix = {arXiv},
       eprint = {0909.1752},
 primaryClass = {astro-ph.EP},
       adsurl = {https://ui.adsabs.harvard.edu/abs/2009ApJ...705..683B}
}

@ARTICLE{2022A&A...658A..75H,
       author = {{Hooton}, M.~J. and {Hoyer}, S. and {Kitzmann}, D. and {Morris}, B.~M. and {Smith}, A.~M.~S. and {Collier Cameron}, A. and {Futyan}, D. and {Maxted}, P.~F.~L. and {Queloz}, D. and {Demory}, B. -O. and {Heng}, K. and {Lendl}, M. and {Cabrera}, J. and {Csizmadia}, Sz. and {Deline}, A. and {Parviainen}, H. and {Salmon}, S. and {Sulis}, S. and {Wilson}, T.~G. and {Bonfanti}, A. and {Brandeker}, A. and {Demangeon}, O.~D.~S. and {Oshagh}, M. and {Persson}, C.~M. and {Scandariato}, G. and {Alibert}, Y. and {Alonso}, R. and {Anglada Escud{\'e}}, G. and {B{\'a}rczy}, T. and {Barrado}, D. and {Barros}, S.~C.~C. and {Baumjohann}, W. and {Beck}, M. and {Beck}, T. and {Benz}, W. and {Billot}, N. and {Bonfils}, X. and {Bourrier}, V. and {Broeg}, C. and {Busch}, M. -D. and {Charnoz}, S. and {Davies}, M.~B. and {Deleuil}, M. and {Delrez}, L. and {Ehrenreich}, D. and {Erikson}, A. and {Farinato}, J. and {Fortier}, A. and {Fossati}, L. and {Fridlund}, M. and {Gandolfi}, D. and {Gillon}, M. and {G{\"u}del}, M. and {Isaak}, K.~G. and {Jones}, K. and {Kiss}, L. and {Laskar}, J. and {Lecavelier des Etangs}, A. and {Lovis}, C. and {Luntzer}, A. and {Magrin}, D. and {Nascimbeni}, V. and {Olofsson}, G. and {Ottensamer}, R. and {Pagano}, I. and {Pall{\'e}}, E. and {Peter}, G. and {Piotto}, G. and {Pollacco}, D. and {Ragazzoni}, R. and {Rando}, N. and {Ratti}, F. and {Rauer}, H. and {Ribas}, I. and {Santos}, N.~C. and {S{\'e}gransan}, D. and {Simon}, A.~E. and {Sousa}, S.~G. and {Steller}, M. and {Szab{\'o}}, Gy. M. and {Thomas}, N. and {Udry}, S. and {Ulmer}, B. and {Van Grootel}, V. and {Walton}, N.~A.},
        title = "{Spi-OPS: Spitzer and CHEOPS confirm the near-polar orbit of MASCARA-1 b and reveal a hint of dayside reflection}",
      journal = {\aap},
         year = 2022,
        month = feb,
       volume = {658},
          eid = {A75},
        pages = {A75},
          doi = {10.1051/0004-6361/202141645},
archivePrefix = {arXiv},
       eprint = {2109.05031},
 primaryClass = {astro-ph.EP},
       adsurl = {https://ui.adsabs.harvard.edu/abs/2022A&A...658A..75H}
}

@ARTICLE{2022A&A...666A.118J,
       author = {{Jones}, K. and {Morris}, B.~M. and {Demory}, B. -O. and {Heng}, K. and {Hooton}, M.~J. and {Billot}, N. and {Ehrenreich}, D. and {Hoyer}, S. and {Simon}, A.~E. and {Lendl}, M. and {Demangeon}, O.~D.~S. and {Sousa}, S.~G. and {Bonfanti}, A. and {Wilson}, T.~G. and {Salmon}, S. and {Csizmadia}, Sz. and {Parviainen}, H. and {Bruno}, G. and {Alibert}, Y. and {Alonso}, R. and {Anglada}, G. and {B{\'a}rczy}, T. and {Barrado}, D. and {Barros}, S.~C.~C. and {Baumjohann}, W. and {Beck}, M. and {Beck}, T. and {Benz}, W. and {Bonfils}, X. and {Brandeker}, A. and {Broeg}, C. and {Cabrera}, J. and {Charnoz}, S. and {Collier Cameron}, A. and {Davies}, M.~B. and {Deleuil}, M. and {Deline}, A. and {Delrez}, L. and {Erikson}, A. and {Fortier}, A. and {Fossati}, L. and {Fridlund}, M. and {Gandolfi}, D. and {Gillon}, M. and {G{\"u}del}, M. and {Isaak}, K.~G. and {Kiss}, L.~L. and {Laskar}, J. and {Lecavelier des Etangs}, A. and {Lovis}, C. and {Magrin}, D. and {Maxted}, P.~F.~L. and {Nascimbeni}, V. and {Olofsson}, G. and {Ottensamer}, R. and {Pagano}, I. and {Pall{\'e}}, E. and {Peter}, G. and {Piotto}, G. and {Pollacco}, D. and {Queloz}, D. and {Ragazzoni}, R. and {Rando}, N. and {Ratti}, F. and {Rauer}, H. and {Reimers}, C. and {Ribas}, I. and {Santos}, N.~C. and {Scandariato}, G. and {S{\'e}gransan}, D. and {Smith}, A.~M.~S. and {Steller}, M. and {Szab{\'o}}, Gy. M. and {Thomas}, N. and {Udry}, S. and {Van Grootel}, V. and {Walter}, I. and {Walton}, N.~A. and {Wang Jungo}, W.},
        title = "{The stable climate of KELT-9b}",
      journal = {\aap},
         year = 2022,
        month = oct,
       volume = {666},
          eid = {A118},
        pages = {A118},
          doi = {10.1051/0004-6361/202243823},
archivePrefix = {arXiv},
       eprint = {2208.04818},
 primaryClass = {astro-ph.EP},
       adsurl = {https://ui.adsabs.harvard.edu/abs/2022A&A...666A.118J}
}

@ARTICLE{2015MNRAS.450.3233P,
       author = {{Parviainen}, Hannu},
        title = "{PYTRANSIT: fast and easy exoplanet transit modelling in PYTHON}",
      journal = {\mnras},
         year = 2015,
        month = jul,
       volume = {450},
       number = {3},
        pages = {3233-3238},
          doi = {10.1093/mnras/stv894},
archivePrefix = {arXiv},
       eprint = {1504.07433},
 primaryClass = {astro-ph.EP},
       adsurl = {https://ui.adsabs.harvard.edu/abs/2015MNRAS.450.3233P}
}

@ARTICLE{2013A&A...553A...6H,
       author = {{Husser}, T. -O. and {Wende-von Berg}, S. and {Dreizler}, S. and {Homeier}, D. and {Reiners}, A. and {Barman}, T. and {Hauschildt}, P.~H.},
        title = "{A new extensive library of PHOENIX stellar atmospheres and synthetic spectra}",
      journal = {\aap},
         year = 2013,
        month = may,
       volume = {553},
          eid = {A6},
        pages = {A6},
          doi = {10.1051/0004-6361/201219058},
archivePrefix = {arXiv},
       eprint = {1303.5632},
 primaryClass = {astro-ph.SR},
       adsurl = {https://ui.adsabs.harvard.edu/abs/2013A&A...553A...6H}
}

@ARTICLE{2013MNRAS.435.2152K,
       author = {{Kipping}, David M.},
        title = "{Efficient, uninformative sampling of limb darkening coefficients for two-parameter laws}",
      journal = {\mnras},
         year = 2013,
        month = nov,
       volume = {435},
       number = {3},
        pages = {2152-2160},
          doi = {10.1093/mnras/stt1435},
archivePrefix = {arXiv},
       eprint = {1308.0009},
 primaryClass = {astro-ph.SR},
       adsurl = {https://ui.adsabs.harvard.edu/abs/2013MNRAS.435.2152K}
}

@ARTICLE{2015ApJ...805...28M,
       author = {{Masuda}, Kento},
        title = "{Spin-Orbit Angles of Kepler-13Ab and HAT-P-7b from Gravity-darkened Transit Light Curves}",
      journal = {\apj},
         year = 2015,
        month = may,
       volume = {805},
       number = {1},
          eid = {28},
        pages = {28},
          doi = {10.1088/0004-637X/805/1/28},
archivePrefix = {arXiv},
       eprint = {1503.05446},
 primaryClass = {astro-ph.EP},
       adsurl = {https://ui.adsabs.harvard.edu/abs/2015ApJ...805...28M}
}

@ARTICLE{2022AJ....163..228P,
       author = {{Patel}, Jayshil A. and {Espinoza}, N{\'e}stor},
        title = "{Empirical Limb-darkening Coefficients and Transit Parameters of Known Exoplanets from TESS}",
      journal = {\aj},
         year = 2022,
        month = may,
       volume = {163},
       number = {5},
          eid = {228},
        pages = {228},
          doi = {10.3847/1538-3881/ac5f55},
archivePrefix = {arXiv},
       eprint = {2203.05661},
 primaryClass = {astro-ph.EP},
       adsurl = {https://ui.adsabs.harvard.edu/abs/2022AJ....163..228P}
}

@ARTICLE{2008ApJ...678L.129C,
       author = {{Cowan}, Nicolas B. and {Agol}, Eric},
        title = "{Inverting Phase Functions to Map Exoplanets}",
      journal = {\apjl},
         year = 2008,
        month = may,
       volume = {678},
       number = {2},
        pages = {L129},
          doi = {10.1086/588553},
archivePrefix = {arXiv},
       eprint = {0803.3622},
 primaryClass = {astro-ph},
       adsurl = {https://ui.adsabs.harvard.edu/abs/2008ApJ...678L.129C}
}

@ARTICLE{2023Natur.620...67K,
       author = {{Kempton}, Eliza M. -R. and {Zhang}, Michael and {Bean}, Jacob L. and {Steinrueck}, Maria E. and {Piette}, Anjali A.~A. and {Parmentier}, Vivien and {Malsky}, Isaac and {Roman}, Michael T. and {Rauscher}, Emily and {Gao}, Peter and {Bell}, Taylor J. and {Xue}, Qiao and {Taylor}, Jake and {Savel}, Arjun B. and {Arnold}, Kenneth E. and {Nixon}, Matthew C. and {Stevenson}, Kevin B. and {Mansfield}, Megan and {Kendrew}, Sarah and {Zieba}, Sebastian and {Ducrot}, Elsa and {Dyrek}, Achr{\`e}ne and {Lagage}, Pierre-Olivier and {Stassun}, Keivan G. and {Henry}, Gregory W. and {Barman}, Travis and {Lupu}, Roxana and {Malik}, Matej and {Kataria}, Tiffany and {Ih}, Jegug and {Fu}, Guangwei and {Welbanks}, Luis and {McGill}, Peter},
        title = "{A reflective, metal-rich atmosphere for GJ 1214b from its JWST phase curve}",
      journal = {\nat},
         year = 2023,
        month = aug,
       volume = {620},
       number = {7972},
        pages = {67-71},
          doi = {10.1038/s41586-023-06159-5},
archivePrefix = {arXiv},
       eprint = {2305.06240},
 primaryClass = {astro-ph.EP},
       adsurl = {https://ui.adsabs.harvard.edu/abs/2023Natur.620...67K}
}

@article{celerite1,
   author = {{Foreman-Mackey}, D. and {Agol}, E. and {Ambikasaran}, S. and
            {Angus}, R.},
    title = "{Fast and Scalable Gaussian Process Modeling with Applications to
              Astronomical Time Series}",
  journal = {\aj},
     year = 2017,
    month = dec,
   volume = 154,
    pages = {220},
      doi = {10.3847/1538-3881/aa9332},
   adsurl = {http://adsabs.harvard.edu/abs/2017AJ....154..220F}
}

@article{celerite2,
   author = {{Foreman-Mackey}, D.},
    title = "{Scalable Backpropagation for Gaussian Processes using Celerite}",
  journal = {Research Notes of the American Astronomical Society},
     year = 2018,
    month = feb,
   volume = 2,
   number = 1,
    pages = {31},
      doi = {10.3847/2515-5172/aaaf6c},
   adsurl = {http://adsabs.harvard.edu/abs/2018RNAAS...2a..31F}
}

@INPROCEEDINGS{2004AIPC..735..395S,
       author = {{Skilling}, John},
        title = "{Nested Sampling}",
    booktitle = {Bayesian Inference and Maximum Entropy Methods in Science and Engineering: 24th International Workshop on Bayesian Inference and Maximum Entropy Methods in Science and Engineering},
         year = 2004,
       editor = {{Fischer}, Rainer and {Preuss}, Roland and {Toussaint}, Udo Von},
       series = {American Institute of Physics Conference Series},
       volume = {735},
        month = nov,
        pages = {395-405},
          doi = {10.1063/1.1835238},
       adsurl = {https://ui.adsabs.harvard.edu/abs/2004AIPC..735..395S}
}

@article{10.1214/06-BA127,
author = {John Skilling},
title = {{Nested sampling for general Bayesian computation}},
volume = {1},
journal = {Bayesian Analysis},
number = {4},
publisher = {International Society for Bayesian Analysis},
pages = {833 -- 859},
year = {2006},
doi = {10.1214/06-BA127},
URL = {https://doi.org/10.1214/06-BA127}
}

@ARTICLE{2019S&C....29..891H,
       author = {{Higson}, Edward and {Handley}, Will and {Hobson}, Mike and {Lasenby}, Anthony},
        title = "{Dynamic nested sampling: an improved algorithm for parameter estimation and evidence calculation}",
      journal = {Statistics and Computing},
         year = 2019,
        month = sep,
       volume = {29},
       number = {5},
        pages = {891-913},
          doi = {10.1007/s11222-018-9844-0},
archivePrefix = {arXiv},
       eprint = {1704.03459},
 primaryClass = {stat.CO},
       adsurl = {https://ui.adsabs.harvard.edu/abs/2019S&C....29..891H}
}

@ARTICLE{2020MNRAS.493.3132S,
       author = {{Speagle}, Joshua S.},
        title = "{DYNESTY: a dynamic nested sampling package for estimating Bayesian posteriors and evidences}",
      journal = {\mnras},
         year = 2020,
        month = apr,
       volume = {493},
       number = {3},
        pages = {3132-3158},
          doi = {10.1093/mnras/staa278},
archivePrefix = {arXiv},
       eprint = {1904.02180},
 primaryClass = {astro-ph.IM},
       adsurl = {https://ui.adsabs.harvard.edu/abs/2020MNRAS.493.3132S}
}

@ARTICLE{2019NatAs...3.1092K,
       author = {{Keating}, Dylan and {Cowan}, Nicolas B. and {Dang}, Lisa},
        title = "{Uniformly hot nightside temperatures on short-period gas giants}",
      journal = {Nature Astronomy},
         year = 2019,
        month = aug,
       volume = {3},
        pages = {1092-1098},
          doi = {10.1038/s41550-019-0859-z},
archivePrefix = {arXiv},
       eprint = {1809.00002},
 primaryClass = {astro-ph.EP},
       adsurl = {https://ui.adsabs.harvard.edu/abs/2019NatAs...3.1092K}
}

@ARTICLE{2022A&A...660A.123M,
       author = {{Morris}, Brett M. and {Heng}, Kevin and {Jones}, Kathryn and {Piaulet}, Caroline and {Demory}, Brice-Olivier and {Kitzmann}, Daniel and {Jens Hoeijmakers}, H.},
        title = "{Physically-motivated basis functions for temperature maps of exoplanets}",
      journal = {\aap},
         year = 2022,
        month = apr,
       volume = {660},
          eid = {A123},
        pages = {A123},
          doi = {10.1051/0004-6361/202142135},
archivePrefix = {arXiv},
       eprint = {2110.11837},
 primaryClass = {astro-ph.EP},
       adsurl = {https://ui.adsabs.harvard.edu/abs/2022A&A...660A.123M}
}

@ARTICLE{2018NatAs...2..220D,
       author = {{Dang}, Lisa and {Cowan}, Nicolas B. and {Schwartz}, Joel C. and {Rauscher}, Emily and {Zhang}, Michael and {Knutson}, Heather A. and {Line}, Michael and {Dobbs-Dixon}, Ian and {Deming}, Drake and {Sundararajan}, Sudarsan and {Fortney}, Jonathan J. and {Zhao}, Ming},
        title = "{Detection of a westward hotspot offset in the atmosphere of hot gas giant CoRoT-2b}",
      journal = {Nature Astronomy},
         year = 2018,
        month = mar,
       volume = {2},
        pages = {220-227},
          doi = {10.1038/s41550-017-0351-6},
archivePrefix = {arXiv},
       eprint = {1801.06548},
 primaryClass = {astro-ph.EP},
       adsurl = {https://ui.adsabs.harvard.edu/abs/2018NatAs...2..220D}
}

@ARTICLE{2011ApJ...729...54C,
       author = {{Cowan}, Nicolas B. and {Agol}, Eric},
        title = "{The Statistics of Albedo and Heat Recirculation on Hot Exoplanets}",
      journal = {\apj},
         year = 2011,
        month = mar,
       volume = {729},
       number = {1},
          eid = {54},
        pages = {54},
          doi = {10.1088/0004-637X/729/1/54},
archivePrefix = {arXiv},
       eprint = {1001.0012},
 primaryClass = {astro-ph.EP},
       adsurl = {https://ui.adsabs.harvard.edu/abs/2011ApJ...729...54C}
}

@ARTICLE{2022NatAs...6..449P,
       author = {{Prinoth}, Bibiana and {Hoeijmakers}, H. Jens and {Kitzmann}, Daniel and {Sandvik}, Elin and {Seidel}, Julia V. and {Lendl}, Monika and {Borsato}, Nicholas W. and {Thorsbro}, Brian and {Anderson}, David R. and {Barrado}, David and {Kravchenko}, Kateryna and {Allart}, Romain and {Bourrier}, Vincent and {Cegla}, Heather M. and {Ehrenreich}, David and {Fisher}, Chloe and {Lovis}, Christophe and {Guzm{\'a}n-Mesa}, Andrea and {Grimm}, Simon and {Hooton}, Matthew and {Morris}, Brett M. and {Oreshenko}, Maria and {Pino}, Lorenzo and {Heng}, Kevin},
        title = "{Titanium oxide and chemical inhomogeneity in the atmosphere of the exoplanet WASP-189 b}",
      journal = {Nature Astronomy},
         year = 2022,
        month = jan,
       volume = {6},
        pages = {449-457},
          doi = {10.1038/s41550-021-01581-z},
archivePrefix = {arXiv},
       eprint = {2111.12732},
 primaryClass = {astro-ph.EP},
       adsurl = {https://ui.adsabs.harvard.edu/abs/2022NatAs...6..449P}
}

@ARTICLE{2023A&A...678A.182P,
       author = {{Prinoth}, B. and {Hoeijmakers}, H.~J. and {Pelletier}, S. and {Kitzmann}, D. and {Morris}, B.~M. and {Seifahrt}, A. and {Kasper}, D. and {Korhonen}, H.~H. and {Burheim}, M. and {Bean}, J.~L. and {Benneke}, B. and {Borsato}, N.~W. and {Brady}, M. and {Grimm}, S.~L. and {Luque}, R. and {St{\"u}rmer}, J. and {Thorsbro}, B.},
        title = "{Time-resolved transmission spectroscopy of the ultra-hot Jupiter WASP-189 b}",
      journal = {\aap},
         year = 2023,
        month = oct,
       volume = {678},
          eid = {A182},
        pages = {A182},
          doi = {10.1051/0004-6361/202347262},
archivePrefix = {arXiv},
       eprint = {2308.04523},
 primaryClass = {astro-ph.EP},
       adsurl = {https://ui.adsabs.harvard.edu/abs/2023A&A...678A.182P}
}

@ARTICLE{2025A&A...693A..72L,
       author = {{Lesjak}, F. and {Nortmann}, L. and {Cont}, D. and {Yan}, F. and {Reiners}, A. and {Piskunov}, N. and {Hatzes}, A. and {Boldt-Christmas}, L. and {Czesla}, S. and {Lavail}, A. and {Nagel}, E. and {Rains}, A.~D. and {Rengel}, M. and {Seemann}, U. and {Shulyak}, D.},
        title = "{Retrieving wind properties from the ultra-hot dayside of WASP-189 b with CRIRES$^{+}$}",
      journal = {\aap},
         year = 2025,
        month = jan,
       volume = {693},
          eid = {A72},
        pages = {A72},
          doi = {10.1051/0004-6361/202451391},
archivePrefix = {arXiv},
       eprint = {2411.19662},
 primaryClass = {astro-ph.EP},
       adsurl = {https://ui.adsabs.harvard.edu/abs/2025A&A...693A..72L}
}

@ARTICLE{2024A&A...683A...1S,
       author = {{Singh}, V. and {Scandariato}, G. and {Smith}, A.~M.~S. and {Cubillos}, P.~E. and {Lendl}, M. and {Billot}, N. and {Fortier}, A. and {Queloz}, D. and {Sousa}, S.~G. and {Csizmadia}, Sz. and {Brandeker}, A. and {Carone}, L. and {Wilson}, T.~G. and {Akinsanmi}, B. and {Patel}, J.~A. and {Krenn}, A. and {Demangeon}, O.~D.~S. and {Bruno}, G. and {Pagano}, I. and {Hooton}, M.~J. and {Cabrera}, J. and {Santos}, N.~C. and {Alibert}, Y. and {Alonso}, R. and {Asquier}, J. and {B{\'a}rczy}, T. and {Navascues}, D. Barrado and {Barros}, S.~C.~C. and {Baumjohann}, W. and {Beck}, M. and {Beck}, T. and {Benz}, W. and {Bergomi}, M. and {Bonfanti}, A. and {Bonfils}, X. and {Borsato}, L. and {Broeg}, C. and {Charnoz}, S. and {Cameron}, A. Collier and {Davies}, M.~B. and {Deleuil}, M. and {Deline}, A. and {Delrez}, L. and {Demory}, B. -O. and {Ehrenreich}, D. and {Erikson}, A. and {Fossati}, L. and {Fridlund}, M. and {Gandolfi}, D. and {Gillon}, M. and {G{\"u}del}, M. and {G{\"u}nther}, M.~N. and {Harre}, J. -V. and {Heitzmann}, A. and {Helling}, Ch. and {Hoyer}, S. and {Isaak}, K.~G. and {Kiss}, L.~L. and {Lam}, K.~W.~F. and {Laskar}, J. and {des Etangs}, A. Lecavelier and {Magrin}, D. and {Maxted}, P.~F.~L. and {Mischler}, H. and {Mordasini}, C. and {Nascimbeni}, V. and {Olofsson}, G. and {Ottensamer}, R. and {Pall{\'e}}, E. and {Peter}, G. and {Piotto}, G. and {Pollacco}, D. and {Ragazzoni}, R. and {Rando}, N. and {Rauer}, H. and {Ribas}, I. and {Salmon}, S. and {S{\'e}gransan}, D. and {Simon}, A.~E. and {Stalport}, M. and {Steinberger}, M. and {Szab{\'o}}, Gy. M. and {Thomas}, N. and {Udry}, S. and {Ulmer}, B. and {Van Grootel}, V. and {Venturini}, J. and {Villaver}, E. and {Walton}, N.~A. and {Zingales}, T.},
        title = "{CHEOPS observations of KELT-20 b/MASCARA-2 b: An aligned orbit and signs of variability from a reflective day side}",
      journal = {\aap},
         year = 2024,
        month = mar,
       volume = {683},
          eid = {A1},
        pages = {A1},
          doi = {10.1051/0004-6361/202347533},
archivePrefix = {arXiv},
       eprint = {2311.03264},
 primaryClass = {astro-ph.EP},
       adsurl = {https://ui.adsabs.harvard.edu/abs/2024A&A...683A...1S}
}

@ARTICLE{2015ApJ...805..132D,
       author = {{Deming}, Drake and {Knutson}, Heather and {Kammer}, Joshua and {Fulton}, Benjamin J. and {Ingalls}, James and {Carey}, Sean and {Burrows}, Adam and {Fortney}, Jonathan J. and {Todorov}, Kamen and {Agol}, Eric and {Cowan}, Nicolas and {Desert}, Jean-Michel and {Fraine}, Jonathan and {Langton}, Jonathan and {Morley}, Caroline and {Showman}, Adam P.},
        title = "{Spitzer Secondary Eclipses of the Dense, Modestly-irradiated, Giant Exoplanet HAT-P-20b Using Pixel-level Decorrelation}",
      journal = {\apj},
         year = 2015,
        month = jun,
       volume = {805},
       number = {2},
          eid = {132},
        pages = {132},
          doi = {10.1088/0004-637X/805/2/132},
archivePrefix = {arXiv},
       eprint = {1411.7404},
 primaryClass = {astro-ph.EP},
       adsurl = {https://ui.adsabs.harvard.edu/abs/2015ApJ...805..132D}
}

@ARTICLE{2016AJ....152..100L,
       author = {{Luger}, Rodrigo and {Agol}, Eric and {Kruse}, Ethan and {Barnes}, Rory and {Becker}, Andrew and {Foreman-Mackey}, Daniel and {Deming}, Drake},
        title = "{EVEREST: Pixel Level Decorrelation of K2 Light Curves}",
      journal = {\aj},
         year = 2016,
        month = oct,
       volume = {152},
       number = {4},
          eid = {100},
        pages = {100},
          doi = {10.3847/0004-6256/152/4/100},
archivePrefix = {arXiv},
       eprint = {1607.00524},
 primaryClass = {astro-ph.EP},
       adsurl = {https://ui.adsabs.harvard.edu/abs/2016AJ....152..100L}
}

@ARTICLE{2024A&A...690A.159P,
       author = {{Patel}, J.~A. and {Brandeker}, A. and {Kitzmann}, D. and {Petit dit de la Roche}, D.~J.~M. and {Bello-Arufe}, A. and {Heng}, K. and {Meier Vald{\'e}s}, E. and {Persson}, C.~M. and {Zhang}, M. and {Demory}, B. -O. and {Bourrier}, V. and {Deline}, A. and {Ehrenreich}, D. and {Fridlund}, M. and {Hu}, R. and {Lendl}, M. and {Oza}, A.~V. and {Alibert}, Y. and {Hooton}, M.~J.},
        title = "{JWST reveals the rapid and strong day-side variability of 55 Cancri e}",
      journal = {\aap},
         year = 2024,
        month = oct,
       volume = {690},
          eid = {A159},
        pages = {A159},
          doi = {10.1051/0004-6361/202450748},
archivePrefix = {arXiv},
       eprint = {2407.12898},
 primaryClass = {astro-ph.EP},
       adsurl = {https://ui.adsabs.harvard.edu/abs/2024A&A...690A.159P}
}

@ARTICLE{2014Sci...346..838S,
       author = {{Stevenson}, Kevin B. and {D{\'e}sert}, Jean-Michel and {Line}, Michael R. and {Bean}, Jacob L. and {Fortney}, Jonathan J. and {Showman}, Adam P. and {Kataria}, Tiffany and {Kreidberg}, Laura and {McCullough}, Peter R. and {Henry}, Gregory W. and {Charbonneau}, David and {Burrows}, Adam and {Seager}, Sara and {Madhusudhan}, Nikku and {Williamson}, Michael H. and {Homeier}, Derek},
        title = "{Thermal structure of an exoplanet atmosphere from phase-resolved emission spectroscopy}",
      journal = {Science},
         year = 2014,
        month = nov,
       volume = {346},
       number = {6211},
        pages = {838-841},
          doi = {10.1126/science.1256758},
archivePrefix = {arXiv},
       eprint = {1410.2241},
 primaryClass = {astro-ph.EP},
       adsurl = {https://ui.adsabs.harvard.edu/abs/2014Sci...346..838S}
}

@ARTICLE{2018AJ....156...10M,
       author = {{Mansfield}, Megan and {Bean}, Jacob L. and {Line}, Michael R. and {Parmentier}, Vivien and {Kreidberg}, Laura and {D{\'e}sert}, Jean-Michel and {Fortney}, Jonathan J. and {Stevenson}, Kevin B. and {Arcangeli}, Jacob and {Dragomir}, Diana},
        title = "{An HST/WFC3 Thermal Emission Spectrum of the Hot Jupiter HAT-P-7b}",
      journal = {\aj},
         year = 2018,
        month = jul,
       volume = {156},
       number = {1},
          eid = {10},
        pages = {10},
          doi = {10.3847/1538-3881/aac497},
archivePrefix = {arXiv},
       eprint = {1805.00424},
 primaryClass = {astro-ph.EP},
       adsurl = {https://ui.adsabs.harvard.edu/abs/2018AJ....156...10M}
}

@ARTICLE{2017ApJ...850L..32S,
       author = {{Sheppard}, Kyle B. and {Mandell}, Avi M. and {Tamburo}, Patrick and {Gandhi}, Siddharth and {Pinhas}, Arazi and {Madhusudhan}, Nikku and {Deming}, Drake},
        title = "{Evidence for a Dayside Thermal Inversion and High Metallicity for the Hot Jupiter WASP-18b}",
      journal = {\apjl},
         year = 2017,
        month = dec,
       volume = {850},
       number = {2},
          eid = {L32},
        pages = {L32},
          doi = {10.3847/2041-8213/aa9ae9},
archivePrefix = {arXiv},
       eprint = {1711.10491},
 primaryClass = {astro-ph.EP},
       adsurl = {https://ui.adsabs.harvard.edu/abs/2017ApJ...850L..32S}
}

@ARTICLE{2021NatAs...5.1224M,
       author = {{Mansfield}, Megan and {Line}, Michael R. and {Bean}, Jacob L. and {Fortney}, Jonathan J. and {Parmentier}, Vivien and {Wiser}, Lindsey and {Kempton}, Eliza M. -R. and {Gharib-Nezhad}, Ehsan and {Sing}, David K. and {L{\'o}pez-Morales}, Mercedes and {Baxter}, Claire and {D{\'e}sert}, Jean-Michel and {Swain}, Mark R. and {Roudier}, Gael M.},
        title = "{A unique hot Jupiter spectral sequence with evidence for compositional diversity}",
      journal = {Nature Astronomy},
         year = 2021,
        month = dec,
       volume = {5},
        pages = {1224-1232},
          doi = {10.1038/s41550-021-01455-4},
archivePrefix = {arXiv},
       eprint = {2110.11272},
 primaryClass = {astro-ph.EP},
       adsurl = {https://ui.adsabs.harvard.edu/abs/2021NatAs...5.1224M}
}

@ARTICLE{2018ApJ...855L..30A,
       author = {{Arcangeli}, Jacob and {D{\'e}sert}, Jean-Michel and {Line}, Michael R. and {Bean}, Jacob L. and {Parmentier}, Vivien and {Stevenson}, Kevin B. and {Kreidberg}, Laura and {Fortney}, Jonathan J. and {Mansfield}, Megan and {Showman}, Adam P.},
        title = "{H$^{-}$ Opacity and Water Dissociation in the Dayside Atmosphere of the Very Hot Gas Giant WASP-18b}",
      journal = {\apjl},
         year = 2018,
        month = mar,
       volume = {855},
       number = {2},
          eid = {L30},
        pages = {L30},
          doi = {10.3847/2041-8213/aab272},
archivePrefix = {arXiv},
       eprint = {1801.02489},
 primaryClass = {astro-ph.EP},
       adsurl = {https://ui.adsabs.harvard.edu/abs/2018ApJ...855L..30A}
}

@ARTICLE{2018A&A...617A.110P,
       author = {{Parmentier}, Vivien and {Line}, Mike R. and {Bean}, Jacob L. and {Mansfield}, Megan and {Kreidberg}, Laura and {Lupu}, Roxana and {Visscher}, Channon and {D{\'e}sert}, Jean-Michel and {Fortney}, Jonathan J. and {Deleuil}, Magalie and {Arcangeli}, Jacob and {Showman}, Adam P. and {Marley}, Mark S.},
        title = "{From thermal dissociation to condensation in the atmospheres of ultra hot Jupiters: WASP-121b in context}",
      journal = {\aap},
         year = 2018,
        month = sep,
       volume = {617},
          eid = {A110},
        pages = {A110},
          doi = {10.1051/0004-6361/201833059},
archivePrefix = {arXiv},
       eprint = {1805.00096},
 primaryClass = {astro-ph.EP},
       adsurl = {https://ui.adsabs.harvard.edu/abs/2018A&A...617A.110P}
}

@ARTICLE{2023Natur.620..292C,
       author = {{Coulombe}, Louis-Philippe and {Benneke}, Bj{\"o}rn and {Challener}, Ryan and {Piette}, Anjali A.~A. and {Wiser}, Lindsey S. and {Mansfield}, Megan and {MacDonald}, Ryan J. and {Beltz}, Hayley and {Feinstein}, Adina D. and {Radica}, Michael and {Savel}, Arjun B. and {Dos Santos}, Leonardo A. and {Bean}, Jacob L. and {Parmentier}, Vivien and {Wong}, Ian and {Rauscher}, Emily and {Komacek}, Thaddeus D. and {Kempton}, Eliza M. -R. and {Tan}, Xianyu and {Hammond}, Mark and {Lewis}, Neil T. and {Line}, Michael R. and {Lee}, Elspeth K.~H. and {Shivkumar}, Hinna and {Crossfield}, Ian J.~M. and {Nixon}, Matthew C. and {Rackham}, Benjamin V. and {Wakeford}, Hannah R. and {Welbanks}, Luis and {Zhang}, Xi and {Batalha}, Natalie M. and {Berta-Thompson}, Zachory K. and {Changeat}, Quentin and {D{\'e}sert}, Jean-Michel and {Espinoza}, N{\'e}stor and {Goyal}, Jayesh M. and {Harrington}, Joseph and {Knutson}, Heather A. and {Kreidberg}, Laura and {L{\'o}pez-Morales}, Mercedes and {Shporer}, Avi and {Sing}, David K. and {Stevenson}, Kevin B. and {Aggarwal}, Keshav and {Ahrer}, Eva-Maria and {Alam}, Munazza K. and {Bell}, Taylor J. and {Blecic}, Jasmina and {Caceres}, Claudio and {Carter}, Aarynn L. and {Casewell}, Sarah L. and {Crouzet}, Nicolas and {Cubillos}, Patricio E. and {Decin}, Leen and {Fortney}, Jonathan J. and {Gibson}, Neale P. and {Heng}, Kevin and {Henning}, Thomas and {Iro}, Nicolas and {Kendrew}, Sarah and {Lagage}, Pierre-Olivier and {Leconte}, J{\'e}r{\'e}my and {Lendl}, Monika and {Lothringer}, Joshua D. and {Mancini}, Luigi and {Mikal-Evans}, Thomas and {Molaverdikhani}, Karan and {Nikolov}, Nikolay K. and {Ohno}, Kazumasa and {Palle}, Enric and {Piaulet}, Caroline and {Redfield}, Seth and {Roy}, Pierre-Alexis and {Tsai}, Shang-Min and {Venot}, Olivia and {Wheatley}, Peter J.},
        title = "{A broadband thermal emission spectrum of the ultra-hot Jupiter WASP-18b}",
      journal = {\nat},
         year = 2023,
        month = aug,
       volume = {620},
       number = {7973},
        pages = {292-298},
          doi = {10.1038/s41586-023-06230-1},
archivePrefix = {arXiv},
       eprint = {2301.08192},
 primaryClass = {astro-ph.EP},
       adsurl = {https://ui.adsabs.harvard.edu/abs/2023Natur.620..292C}
}

@ARTICLE{2015AREPS..43..509H,
       author = {{Heng}, Kevin and {Showman}, Adam P.},
        title = "{Atmospheric Dynamics of Hot Exoplanets}",
      journal = {Annual Review of Earth and Planetary Sciences},
         year = 2015,
        month = may,
       volume = {43},
        pages = {509-540},
          doi = {10.1146/annurev-earth-060614-105146},
archivePrefix = {arXiv},
       eprint = {1407.4150},
 primaryClass = {astro-ph.EP},
       adsurl = {https://ui.adsabs.harvard.edu/abs/2015AREPS..43..509H}
}

@ARTICLE{2016ApJ...821...16K,
       author = {{Komacek}, Thaddeus D. and {Showman}, Adam P.},
        title = "{Atmospheric Circulation of Hot Jupiters: Dayside-Nightside Temperature Differences}",
      journal = {\apj},
         year = 2016,
        month = apr,
       volume = {821},
       number = {1},
          eid = {16},
        pages = {16},
          doi = {10.3847/0004-637X/821/1/16},
archivePrefix = {arXiv},
       eprint = {1601.00069},
 primaryClass = {astro-ph.EP},
       adsurl = {https://ui.adsabs.harvard.edu/abs/2016ApJ...821...16K}
}

@ARTICLE{2018ApJ...857L..20B,
       author = {{Bell}, Taylor J. and {Cowan}, Nicolas B.},
        title = "{Increased Heat Transport in Ultra-hot Jupiter Atmospheres through H$_{2}$ Dissociation and Recombination}",
      journal = {\apjl},
         year = 2018,
        month = apr,
       volume = {857},
       number = {2},
          eid = {L20},
        pages = {L20},
          doi = {10.3847/2041-8213/aabcc8},
archivePrefix = {arXiv},
       eprint = {1802.07725},
 primaryClass = {astro-ph.EP},
       adsurl = {https://ui.adsabs.harvard.edu/abs/2018ApJ...857L..20B}
}

@ARTICLE{2020ApJ...888L..15M,
       author = {{Mansfield}, Megan and {Bean}, Jacob L. and {Stevenson}, Kevin B. and {Komacek}, Thaddeus D. and {Bell}, Taylor J. and {Tan}, Xianyu and {Malik}, Matej and {Beatty}, Thomas G. and {Wong}, Ian and {Cowan}, Nicolas B. and {Dang}, Lisa and {D{\'e}sert}, Jean-Michel and {Fortney}, Jonathan J. and {Gaudi}, B. Scott and {Keating}, Dylan and {Kempton}, Eliza M. -R. and {Kreidberg}, Laura and {Line}, Michael R. and {Parmentier}, Vivien and {Stassun}, Keivan G. and {Swain}, Mark R. and {Zellem}, Robert T.},
        title = "{Evidence for H2 Dissociation and Recombination Heat Transport in the Atmosphere of KELT-9b}",
      journal = {\apjl},
         year = 2020,
        month = jan,
       volume = {888},
       number = {2},
          eid = {L15},
        pages = {L15},
          doi = {10.3847/2041-8213/ab5b09},
archivePrefix = {arXiv},
       eprint = {1910.01567},
 primaryClass = {astro-ph.EP},
       adsurl = {https://ui.adsabs.harvard.edu/abs/2020ApJ...888L..15M}
}

@ARTICLE{2024A&A...685A..60P,
       author = {{Prinoth}, B. and {Hoeijmakers}, H.~J. and {Morris}, B.~M. and {Lam}, M. and {Kitzmann}, D. and {Sedaghati}, E. and {Seidel}, J.~V. and {Lee}, E.~K.~H. and {Thorsbro}, B. and {Borsato}, N.~W. and {Damasceno}, Y.~C. and {Pelletier}, S. and {Seifahrt}, A.},
        title = "{An atlas of resolved spectral features in the transmission spectrum of WASP-189 b with MAROON-X}",
      journal = {\aap},
         year = 2024,
        month = may,
       volume = {685},
          eid = {A60},
        pages = {A60},
          doi = {10.1051/0004-6361/202349125},
archivePrefix = {arXiv},
       eprint = {2403.08863},
 primaryClass = {astro-ph.EP},
       adsurl = {https://ui.adsabs.harvard.edu/abs/2024A&A...685A..60P}
}

@ARTICLE{2023ApJ...954L..23S,
       author = {{Sreejith}, A.~G. and {France}, Kevin and {Fossati}, Luca and {Koskinen}, Tommi T. and {Egan}, Arika and {Cauley}, P. Wilson and {Cubillos}, Patricio. E. and {Ambily}, S. and {Huang}, Chenliang and {Lavvas}, Panayotis and {Fleming}, Brian T. and {Desert}, Jean-Michel and {Nell}, Nicholas and {Petit}, Pascal and {Vidotto}, Aline},
        title = "{CUTE Reveals Escaping Metals in the Upper Atmosphere of the Ultrahot Jupiter WASP-189b}",
      journal = {\apjl},
         year = 2023,
        month = sep,
       volume = {954},
       number = {1},
          eid = {L23},
        pages = {L23},
          doi = {10.3847/2041-8213/acef1c},
archivePrefix = {arXiv},
       eprint = {2308.05726},
 primaryClass = {astro-ph.EP},
       adsurl = {https://ui.adsabs.harvard.edu/abs/2023ApJ...954L..23S}
}

@ARTICLE{2022A&A...661L...6Y,
       author = {{Yan}, F. and {Pall{\'e}}, E. and {Reiners}, A. and {Casasayas-Barris}, N. and {Cont}, D. and {Stangret}, M. and {Nortmann}, L. and {Molli{\`e}re}, P. and {Henning}, Th. and {Chen}, G. and {Molaverdikhani}, K.},
        title = "{Detection of CO emission lines in the dayside atmospheres of WASP-33b and WASP-189b with GIANO}",
      journal = {\aap},
         year = 2022,
        month = may,
       volume = {661},
          eid = {L6},
        pages = {L6},
          doi = {10.1051/0004-6361/202243503},
archivePrefix = {arXiv},
       eprint = {2204.10158},
 primaryClass = {astro-ph.EP},
       adsurl = {https://ui.adsabs.harvard.edu/abs/2022A&A...661L...6Y}
}

@ARTICLE{2024AJ....168..148D,
       author = {{Deibert}, Emily K. and {Langeveld}, Adam B. and {Young}, Mitchell E. and {Flagg}, Laura and {Turner}, Jake D. and {Smith}, Peter C.~B. and {de Mooij}, Ernst J.~W. and {Jayawardhana}, Ray and {Chiboucas}, Kristin and {Gamen}, Roberto and {Hayes}, Christian R. and {Heo}, Jeong-Eun and {Jeong}, Miji and {Kalari}, Venu and {Martioli}, Eder and {Placco}, Vinicius M. and {Xu}, Siyi and {Diaz}, Ruben and {Gomez-Jimenez}, Manuel and {Quiroz}, Carlos and {Ruiz-Carmona}, Roque and {Simpson}, Chris and {McConnachie}, Alan W. and {Pazder}, John and {Burley}, Gregory and {Ireland}, Michael and {Waller}, Fletcher and {Berg}, Trystyn A.~M. and {Robertson}, J. Gordon and {Jones}, David O. and {Labrie}, Kathleen and {Ridgway}, Susan and {Thomas-Osip}, Joanna},
        title = "{High-resolution Dayside Spectroscopy of WASP-189 b: Detection of Iron during the GHOST/Gemini South System Verification Run}",
      journal = {\aj},
         year = 2024,
        month = oct,
       volume = {168},
       number = {4},
          eid = {148},
        pages = {148},
          doi = {10.3847/1538-3881/ad643f},
archivePrefix = {arXiv},
       eprint = {2407.11281},
 primaryClass = {astro-ph.EP},
       adsurl = {https://ui.adsabs.harvard.edu/abs/2024AJ....168..148D}
}

@ARTICLE{2025A&A...700A...9V,
       author = {{Vaulato}, Valentina and {Pelletier}, Stefan and {Ehrenreich}, David and {Allart}, Romain and {Cristo}, Eduardo and {Steiner}, Michal and {Dumusque}, Xavier and {Chakraborty}, Hritam and {Lendl}, Monika and {Srivastava}, Avidaan and {Artigau}, {\'E}tienne and {Baron}, Fr{\'e}d{\'e}rique and {Barros}, Susana C.~C. and {Benneke}, Bj{\"o}rn and {Bonfils}, Xavier and {Bouchy}, Fran{\c{c}}ois and {Bryan}, Marta and {Martins}, Bruno L. Canto and {Cloutier}, Ryan and {Cook}, Neil J. and {Cowan}, Nicolas B. and {De Medeiros}, Jose Renan and {Delfosse}, Xavier and {Doyon}, Ren{\'e} and {Gonz{\'a}lez Hern{\'a}ndez}, Jonay I. and {Lafreni{\`e}re}, David and {de Castro Le{\~a}o}, Izan and {Lovis}, Christophe and {Malo}, Lison and {Melo}, Claudio and {Mignon}, Lucile and {Mordasini}, Christoph and {Pepe}, Francesco and {Rebolo}, Rafael and {Rowe}, Jason and {Santos}, Nuno C. and {S{\'e}gransan}, Damien and {Su{\'a}rez Mascare{\~n}o}, Alejandro and {Udry}, St{\'e}phane and {Valencia}, Diana and {Wade}, Gregg and {Al Moulla}, Khaled and {Almenara}, Jose Manuel and {Akinsanmi}, Babatunde and {Bazinet}, Luc and {Bourrier}, Vincent and {Cadieux}, Charles and {Carmona}, Andres and {Carteret}, Yann and {Silva}, Ana Rita Costa and {Darveau-Bernier}, Antoine and {Dauplaise}, Laurie and {de Lima Gomes}, Roseane and {Delisle}, Jean-Baptiste and {Forveille}, Thierry and {Frensch}, Yolanda and {Gagn{\'e}}, Jonathan and {Genest}, Fr{\'e}d{\'e}ric and {da Silva}, Jo{\~a}o Gomes and {Grieves}, Nolan and {Hobson}, Melissa J. and {Krishnamurthy}, Vigneshwaran and {L'Heureux}, Alexandrine and {Lamontagne}, Pierrot and {Larue}, Pierre and {Lim}, Olivia and {Lo Curto}, Gaspare and {Messias}, Yuri S. and {Moranta}, Leslie and {Mounzer}, Dany and {Nari}, Nicola and {Osborn}, Ares and {Parc}, L{\'e}na and {Piaulet}, Caroline and {Plotnykov}, Mykhaylo and {Psaridi}, Angelica and {Stefanov}, Atanas K. and {Teixeira}, M{\'a}rcio A. and {Vandal}, Thomas and {Wardenier}, Joost P. and {Weisserman}, Drew and {Yariv}, Vincent},
        title = "{Hydride ion continuum hides absorption signatures in the NIRPS near-infrared transmission spectrum of the ultra-hot gas giant WASP-189b}",
      journal = {\aap},
         year = 2025,
        month = aug,
       volume = {700},
          eid = {A9},
        pages = {A9},
          doi = {10.1051/0004-6361/202452972},
archivePrefix = {arXiv},
       eprint = {2507.21229},
 primaryClass = {astro-ph.EP},
       adsurl = {https://ui.adsabs.harvard.edu/abs/2025A&A...700A...9V}
}

@INPROCEEDINGS{2014SPIE.9143E..20R,
       author = {{Ricker}, George R. and {Winn}, Joshua N. and {Vanderspek}, Roland and {Latham}, David W. and {Bakos}, G{\'a}sp{\'a}r. {\'A}. and {Bean}, Jacob L. and {Berta-Thompson}, Zachory K. and {Brown}, Timothy M. and {Buchhave}, Lars and {Butler}, Nathaniel R. and {Butler}, R. Paul and {Chaplin}, William J. and {Charbonneau}, David and {Christensen-Dalsgaard}, J{\o}rgen and {Clampin}, Mark and {Deming}, Drake and {Doty}, John and {De Lee}, Nathan and {Dressing}, Courtney and {Dunham}, E.~W. and {Endl}, Michael and {Fressin}, Francois and {Ge}, Jian and {Henning}, Thomas and {Holman}, Matthew J. and {Howard}, Andrew W. and {Ida}, Shigeru and {Jenkins}, Jon and {Jernigan}, Garrett and {Johnson}, John A. and {Kaltenegger}, Lisa and {Kawai}, Nobuyuki and {Kjeldsen}, Hans and {Laughlin}, Gregory and {Levine}, Alan M. and {Lin}, Douglas and {Lissauer}, Jack J. and {MacQueen}, Phillip and {Marcy}, Geoffrey and {McCullough}, P.~R. and {Morton}, Timothy D. and {Narita}, Norio and {Paegert}, Martin and {Palle}, Enric and {Pepe}, Francesco and {Pepper}, Joshua and {Quirrenbach}, Andreas and {Rinehart}, S.~A. and {Sasselov}, Dimitar and {Sato}, Bun'ei and {Seager}, Sara and {Sozzetti}, Alessandro and {Stassun}, Keivan G. and {Sullivan}, Peter and {Szentgyorgyi}, Andrew and {Torres}, Guillermo and {Udry}, Stephane and {Villasenor}, Joel},
        title = "{Transiting Exoplanet Survey Satellite (TESS)}",
    booktitle = {Space Telescopes and Instrumentation 2014: Optical, Infrared, and Millimeter Wave},
         year = 2014,
       editor = {{Oschmann}, Jacobus M., Jr. and {Clampin}, Mark and {Fazio}, Giovanni G. and {MacEwen}, Howard A.},
       series = {Society of Photo-Optical Instrumentation Engineers (SPIE) Conference Series},
       volume = {9143},
        month = aug,
          eid = {914320},
        pages = {914320},
          doi = {10.1117/12.2063489},
archivePrefix = {arXiv},
       eprint = {1406.0151},
 primaryClass = {astro-ph.EP},
       adsurl = {https://ui.adsabs.harvard.edu/abs/2014SPIE.9143E..20R}
}

@ARTICLE{2012MNRAS.421L.122S,
       author = {{Szab{\'o}}, Gy. M. and {P{\'a}l}, A. and {Derekas}, A. and {Simon}, A.~E. and {Szalai}, T. and {Kiss}, L.~L.},
        title = "{Spin-orbit resonance, transit duration variation and possible secular perturbations in KOI-13}",
      journal = {\mnras},
         year = 2012,
        month = mar,
       volume = {421},
       number = {1},
        pages = {L122-L126},
          doi = {10.1111/j.1745-3933.2012.01219.x},
archivePrefix = {arXiv},
       eprint = {1110.4231},
 primaryClass = {astro-ph.SR},
       adsurl = {https://ui.adsabs.harvard.edu/abs/2012MNRAS.421L.122S}
}

@ARTICLE{2024AJ....168....4H,
       author = {{Hammond}, Mark and {Bell}, Taylor J. and {Challener}, Ryan C. and {Lewis}, Neil T. and {Weiner Mansfield}, Megan and {Malsky}, Isaac and {Rauscher}, Emily and {Bean}, Jacob L. and {Carone}, Ludmila and {Mendon{\c{c}}a}, Jo{\~a}o M. and {Teinturier}, Lucas and {Tan}, Xianyu and {Crouzet}, Nicolas and {Kreidberg}, Laura and {Morello}, Giuseppe and {Parmentier}, Vivien and {Blecic}, Jasmina and {D{\'e}sert}, Jean-Michel and {Helling}, Christiane and {Lagage}, Pierre-Olivier and {Molaverdikhani}, Karan and {Nixon}, Matthew C. and {Rackham}, Benjamin V. and {Yang}, Jingxuan},
        title = "{Two-dimensional Eclipse Mapping of the Hot-Jupiter WASP-43b with JWST MIRI/LRS}",
      journal = {\aj},
         year = 2024,
        month = jul,
       volume = {168},
       number = {1},
          eid = {4},
        pages = {4},
          doi = {10.3847/1538-3881/ad434d},
archivePrefix = {arXiv},
       eprint = {2404.16488},
 primaryClass = {astro-ph.EP},
       adsurl = {https://ui.adsabs.harvard.edu/abs/2024AJ....168....4H}
}

@ARTICLE{Malik2017AJ....153...56M,
       author = {{Malik}, Matej and {Grosheintz}, Luc and {Mendon{\c{c}}a}, Jo{\~a}o M. and {Grimm}, Simon L. and {Lavie}, Baptiste and {Kitzmann}, Daniel and {Tsai}, Shang-Min and {Burrows}, Adam and {Kreidberg}, Laura and {Bedell}, Megan and {Bean}, Jacob L. and {Stevenson}, Kevin B. and {Heng}, Kevin},
        title = "{HELIOS: An Open-source, GPU-accelerated Radiative Transfer Code for Self-consistent Exoplanetary Atmospheres}",
      journal = {\aj},
         year = 2017,
        month = feb,
       volume = {153},
       number = {2},
          eid = {56},
        pages = {56},
          doi = {10.3847/1538-3881/153/2/56},
archivePrefix = {arXiv},
       eprint = {1606.05474},
 primaryClass = {astro-ph.EP},
       adsurl = {https://ui.adsabs.harvard.edu/abs/2017AJ....153...56M}
}

@ARTICLE{Malik2019AJ....157..170M,
       author = {{Malik}, Matej and {Kitzmann}, Daniel and {Mendon{\c{c}}a}, Jo{\~a}o M. and {Grimm}, Simon L. and {Marleau}, Gabriel-Dominique and {Linder}, Esther F. and {Tsai}, Shang-Min and {Heng}, Kevin},
        title = "{Self-luminous and Irradiated Exoplanetary Atmospheres Explored with HELIOS}",
      journal = {\aj},
         year = 2019,
        month = may,
       volume = {157},
       number = {5},
          eid = {170},
        pages = {170},
          doi = {10.3847/1538-3881/ab1084},
archivePrefix = {arXiv},
       eprint = {1903.06794},
 primaryClass = {astro-ph.EP},
       adsurl = {https://ui.adsabs.harvard.edu/abs/2019AJ....157..170M}
}

@article{Stamnes1988ApOpt,
	adsurl = "http://adsabs.harvard.edu/abs/1988ApOpt..27.2502S",
	author = "{Stamnes}, K. and {Tsay}, S.-C. and {Jayaweera}, K. and {Wiscombe}, W.",
	doi = "10.1364/AO.27.002502",
	journal = "\ao",
	month = jun,
	pages = "2502--2509",
	title = "{Numerically stable algorithm for discrete-ordinate-method radiative transfer in multiple scattering and emitting layered media}",
	volume = "27",
	year = "1988"
}

@inproceedings{Hamre2013AIPC.1531..923H,
	adsurl = "http://adsabs.harvard.edu/abs/2013AIPC.1531..923H",
	author = "{Hamre}, B. and {Stamnes}, S. and {Stamnes}, K. and {Stamnes}, J.~J.",
	booktitle = "{American Institute of Physics Conference Series}",
	doi = "10.1063/1.4804922",
	month = may,
	pages = "923--926",
	series = {{Radiation Processes in the Atmosphere and Ocean, American Institute of Physics Conference Series}},
	title = "{C-disort: A versatile tool for radiative transfer in coupled media like the atmosphere-ocean system}",
	volume = 1531,
	year = 2013,
	editor = {{Cahalan}, {R.~F.} and {Fischer}, {J.}}
}

@ARTICLE{Heng2021NatAs...5.1001H,
       author = {{Heng}, Kevin and {Morris}, Brett M. and {Kitzmann}, Daniel},
        title = "{Closed-form ab initio solutions of geometric albedos and reflected light phase curves of exoplanets}",
      journal = {Nature Astronomy},
         year = 2021,
        month = oct,
       volume = {5},
        pages = {1001-1008},
          doi = {10.1038/s41550-021-01444-7},
archivePrefix = {arXiv},
       eprint = {2103.02673},
 primaryClass = {astro-ph.EP},
       adsurl = {https://ui.adsabs.harvard.edu/abs/2021NatAs...5.1001H}
}

@BOOK{Sobolev1975lpsa.book.....S,
       author = {{Sobolev}, V.~V.},
        title = "{Light scattering in planetary atmospheres}",
         year = 1975,
       adsurl = {https://ui.adsabs.harvard.edu/abs/1975lpsa.book.....S}
}

@ARTICLE{Kitzmann2018ApJ...863..183K,
       author = {{Kitzmann}, Daniel and {Heng}, Kevin and {Rimmer}, Paul B. and {Hoeijmakers}, H. Jens and {Tsai}, Shang-Min and {Malik}, Matej and {Lendl}, Monika and {Deitrick}, Russell and {Demory}, Brice-Olivier},
        title = "{The Peculiar Atmospheric Chemistry of KELT-9b}",
      journal = {\apj},
         year = 2018,
        month = aug,
       volume = {863},
       number = {2},
          eid = {183},
        pages = {183},
          doi = {10.3847/1538-4357/aace5a},
archivePrefix = {arXiv},
       eprint = {1804.07137},
 primaryClass = {astro-ph.EP},
       adsurl = {https://ui.adsabs.harvard.edu/abs/2018ApJ...863..183K}
}

@ARTICLE{Lothringer2018ApJ...866...27L,
       author = {{Lothringer}, Joshua D. and {Barman}, Travis and {Koskinen}, Tommi},
        title = "{Extremely Irradiated Hot Jupiters: Non-oxide Inversions, H$^{-}$ Opacity, and Thermal Dissociation of Molecules}",
      journal = {\apj},
         year = 2018,
        month = oct,
       volume = {866},
       number = {1},
          eid = {27},
        pages = {27},
          doi = {10.3847/1538-4357/aadd9e},
archivePrefix = {arXiv},
       eprint = {1805.00038},
 primaryClass = {astro-ph.EP},
       adsurl = {https://ui.adsabs.harvard.edu/abs/2018ApJ...866...27L}
}

@ARTICLE{Stock2018MNRAS.479..865S,
   author = {{Stock}, J.~W. and {Kitzmann}, D. and {Patzer}, A.~B.~C. and
	{Sedlmayr}, E.},
    title = "{FastChem: A computer program for efficient complex chemical equilibrium calculations in the neutral/ionized gas phase with applications to stellar and planetary atmospheres}",
  journal = {\mnras},
archivePrefix = "arXiv",
   eprint = {1804.05010},
 primaryClass = "astro-ph.EP",
     year = 2018,
    month = sep,
   volume = 479,
    pages = {865-874},
      doi = {10.1093/mnras/sty1531},
   adsurl = {http://adsabs.harvard.edu/abs/2018MNRAS.479..865S}
}

@ARTICLE{Stock2022MNRAS.517.4070S,
       author = {{Stock}, Joachim W. and {Kitzmann}, Daniel and {Patzer}, A. Beate C.},
        title = "{FASTCHEM 2 : an improved computer program to determine the gas-phase chemical equilibrium composition for arbitrary element distributions}",
      journal = {\mnras},
         year = 2022,
        month = dec,
       volume = {517},
       number = {3},
        pages = {4070-4080},
          doi = {10.1093/mnras/stac2623},
archivePrefix = {arXiv},
       eprint = {2206.08247},
 primaryClass = {astro-ph.EP},
       adsurl = {https://ui.adsabs.harvard.edu/abs/2022MNRAS.517.4070S}
}

@ARTICLE{Kitzmann2024MNRAS.527.7263K,
       author = {{Kitzmann}, Daniel and {Stock}, Joachim W. and {Patzer}, A. Beate C.},
        title = "{FASTCHEM COND: equilibrium chemistry with condensation and rainout for cool planetary and stellar environments}",
      journal = {\mnras},
         year = 2024,
        month = jan,
       volume = {527},
       number = {3},
        pages = {7263-7283},
          doi = {10.1093/mnras/stad3515},
archivePrefix = {arXiv},
       eprint = {2309.02337},
 primaryClass = {astro-ph.EP},
       adsurl = {https://ui.adsabs.harvard.edu/abs/2024MNRAS.527.7263K}
}

@ARTICLE{2025A&A...699A.150D,
       author = {{Deline}, A. and {Cubillos}, P.~E. and {Carone}, L. and {Demory}, B. -O. and {Lendl}, M. and {Benz}, W. and {Brandeker}, A. and {G{\"u}nther}, M.~N. and {Heitzmann}, A. and {Barros}, S.~C.~C. and {Kreidberg}, L. and {Bruno}, G. and {Kitzmann}, D. and {Bonfanti}, A. and {Farnir}, M. and {Persson}, C.~M. and {Sousa}, S.~G. and {Wilson}, T.~G. and {Ehrenreich}, D. and {Singh}, V. and {Iro}, N. and {Alibert}, Y. and {Alonso}, R. and {B{\'a}rczy}, T. and {Barrado Navascues}, D. and {Baumjohann}, W. and {Bergomi}, M. and {Billot}, N. and {Borsato}, L. and {Broeg}, C. and {Busch}, M. -D. and {Collier Cameron}, A. and {Correia}, A.~C.~M. and {Csizmadia}, Sz. and {Davies}, M.~B. and {Deleuil}, M. and {Delrez}, L. and {Demangeon}, O.~D.~S. and {Derekas}, A. and {Edwards}, B. and {Erikson}, A. and {Fortier}, A. and {Fossati}, L. and {Fridlund}, M. and {Gandolfi}, D. and {Gazeas}, K. and {Gillon}, M. and {G{\"u}del}, M. and {Hasiba}, J. and {Helling}, Ch. and {Isaak}, K.~G. and {Kiss}, L.~L. and {Korth}, J. and {Lam}, K.~W.~F. and {Laskar}, J. and {Lecavelier des {\'E}tangs}, A. and {Magrin}, D. and {Maxted}, P.~F.~L. and {Mer{\'\i}n}, B. and {Mordasini}, C. and {Nascimbeni}, V. and {Olofsson}, G. and {Ottensamer}, R. and {Pagano}, I. and {Pall{\'e}}, E. and {Peter}, G. and {Piazza}, D. and {Piotto}, G. and {Pollacco}, D. and {Queloz}, D. and {Ragazzoni}, R. and {Rando}, N. and {Ratti}, F. and {Rauer}, H. and {Ribas}, I. and {Santos}, N.~C. and {Scandariato}, G. and {S{\'e}gransan}, D. and {Simon}, A.~E. and {Smith}, A.~M.~S. and {Stalport}, M. and {Sulis}, S. and {Szab{\'o}}, Gy. M. and {Udry}, S. and {Van Grootel}, V. and {Venturini}, J. and {Villaver}, E. and {Walton}, N.~A. and {Westerdorff}, K.},
        title = "{Dark skies of the slightly eccentric WASP-18 b from its optical-to-infrared dayside emission}",
      journal = {\aap},
         year = 2025,
        month = jul,
       volume = {699},
          eid = {A150},
        pages = {A150},
          doi = {10.1051/0004-6361/202450939},
archivePrefix = {arXiv},
       eprint = {2505.01544},
 primaryClass = {astro-ph.EP},
       adsurl = {https://ui.adsabs.harvard.edu/abs/2025A&A...699A.150D}
}

@ARTICLE{2015ApJ...810L..23J,
       author = {{Johnson}, Marshall C. and {Cochran}, William D. and {Collier Cameron}, Andrew and {Bayliss}, Daniel},
        title = "{Measurement of the Nodal Precession of WASP-33 b via Doppler Tomography}",
      journal = {\apjl},
         year = 2015,
        month = sep,
       volume = {810},
       number = {2},
          eid = {L23},
        pages = {L23},
          doi = {10.1088/2041-8205/810/2/L23},
archivePrefix = {arXiv},
       eprint = {1508.02398},
 primaryClass = {astro-ph.EP},
       adsurl = {https://ui.adsabs.harvard.edu/abs/2015ApJ...810L..23J}
}

@ARTICLE{2024PASJ...76..374W,
       author = {{Watanabe}, Noriharu and {Narita}, Norio and {Hori}, Yasunori},
        title = "{Nodal precession of a hot Jupiter transiting the edge of a late A-type star TOI-1518}",
      journal = {\pasj},
         year = 2024,
        month = jun,
       volume = {76},
       number = {3},
        pages = {374-385},
          doi = {10.1093/pasj/psae019},
archivePrefix = {arXiv},
       eprint = {2402.17325},
 primaryClass = {astro-ph.EP},
       adsurl = {https://ui.adsabs.harvard.edu/abs/2024PASJ...76..374W}
}

@ARTICLE{2022ApJ...931..111S,
       author = {{Stephan}, Alexander P. and {Wang}, Ji and {Cauley}, P. Wilson and {Gaudi}, B. Scott and {Ilyin}, Ilya and {Johnson}, Marshall C. and {Strassmeier}, Klaus G.},
        title = "{Nodal Precession and Tidal Evolution of Two Hot Jupiters: WASP-33 b and KELT-9 b}",
      journal = {\apj},
         year = 2022,
        month = jun,
       volume = {931},
       number = {2},
          eid = {111},
        pages = {111},
          doi = {10.3847/1538-4357/ac6b9a},
archivePrefix = {arXiv},
       eprint = {2203.02546},
 primaryClass = {astro-ph.EP},
       adsurl = {https://ui.adsabs.harvard.edu/abs/2022ApJ...931..111S}
}

@ARTICLE{2022MNRAS.512.4404W,
       author = {{Watanabe}, Noriharu and {Narita}, Norio and {Palle}, Enric and {Fukui}, Akihiko and {Kusakabe}, Nobuhiko and {Parviainen}, Hannu and {Murgas}, Felipe and {Casasayas-Barris}, N{\'u}ria and {Johnson}, Marshall C. and {Sato}, Bun'ei and {Livingston}, John H. and {de Leon}, Jerome P. and {Mori}, Mayuko and {Nishiumi}, Taku and {Terada}, Yuka and {Esparza-Borges}, Emma and {Kawauchi}, Kiyoe},
        title = "{Nodal precession of WASP-33b for 11 yr by Doppler tomographic and transit photometric observations}",
      journal = {\mnras},
         year = 2022,
        month = may,
       volume = {512},
       number = {3},
        pages = {4404-4418},
          doi = {10.1093/mnras/stac620},
archivePrefix = {arXiv},
       eprint = {2203.02003},
 primaryClass = {astro-ph.EP},
       adsurl = {https://ui.adsabs.harvard.edu/abs/2022MNRAS.512.4404W}
}

\begin{appendix}

\onecolumn

\section{Planetary and stellar parameters}

\begin{table*}[h!]
    \caption{Planetary and stellar parameters used in the light curve analysis}
    \centering
    \begin{tabular}{lcccr}
    \toprule
    \toprule
    Parameters & Symbols & Values & Priors & Units \\
    \midrule
    Planetary parameters &&&&\\
    \quad Orbital period & $P$ & \per & {\small$ \mathcal{N}(2.72402, 3.1828 \times 10^{-5})$} & days \\
    \noalign{\smallskip}
    \quad Transit time & $T_0$ & \Ttra & {\small $\mathcal{N}(2459691.99555, 0.0078935)$} & BJD\\
    \noalign{\smallskip}
    \quad Planet-to-star radius ratio & $R_p/R_\star$ & \rprs & {\small $\mathcal{U}(0.065, 0.080)$} & --\\
    \noalign{\smallskip}
    \quad Impact parameter & $b$ & \bb & {\small $\mathcal{N} (0.433, 0.020)$} & -- \\
    \noalign{\smallskip}
    \quad Scaled semi-major axis & $a/R_\star$ & \ar & {\small $\mathcal{N} (4.587, 0.050)$} & --\\
    \noalign{\smallskip}
    \quad Eccentricity & $e$ & {\small 0} & {\small Fixed} & --\\
    \noalign{\smallskip}
    \quad Argument of periastron passage & $\omega$ & {\small 90} & {\small Fixed} & deg\\
    \noalign{\smallskip}
    \quad Projected orbital obliquity & $\lambda_p$ & \lamp & {\small $\mathcal{N} (91.7, 1.2)$} & deg\\
    \noalign{\smallskip}
    \multirow{5}{*}{Phase curve parameters} & $E$ & \ECowan & {\small $\mathcal{U} (0, 500)$} & ppm \\
    \noalign{\smallskip}
    & $C_1$ & \COneCowan & {\small $\mathcal{U} (-10^6, 10^6)$} & ppm \\
    \noalign{\smallskip}
    & $D_1$ & \DOneCowan & {\small $\mathcal{U} (-10^6, 10^6)$} & ppm \\
    \noalign{\smallskip}
    & $C_2$ & \CTwoCowan & {\small Fixed} & ppm \\
    \noalign{\smallskip}
    & $D_2$ & \DTwoCowan & {\small Fixed} & ppm \\
    \noalign{\smallskip}
    
    \midrule
    
    Stellar parameters &&&& \\
    \quad Stellar mass & $M_\star$ & \Mstar & {\small $\mathcal{N}(2.031, 0.098)$} & M$_\odot$ \\
    \noalign{\smallskip}
    \quad Stellar radius & $R_\star$ & \Rstar & {\small $\mathcal{N}(2.365, 0.025)$} & R$_\odot$ \\
    \noalign{\smallskip}
    \quad Pole temperature & $T_{\text{pole}}$ & \Tpole & {\small $\mathcal{N}(8000, 80)$} & K \\
    \noalign{\smallskip}
    \quad Projected rotation period & $v\sin{i_\star}$ & \vsini & {\small $\mathcal{N}(92.5, 2.5)$} & km/s \\
    \noalign{\smallskip}
    \quad Stellar obliquity & $\phi_\star$ & \phistar & {\small $\mathcal{N}(21.8, 1.6)$} & deg \\
    \noalign{\smallskip}
    \quad Gravity darkening coefficient & $\beta$ & \gdc & {\small Fixed} & -- \\
    \noalign{\smallskip}
    \quad Limb darkening coefficients &&&& \\
    \multirow{2}{*}{\qquad \citet{2013MNRAS.435.2152K} parametrisation} & $q_{1_\mathrm{TESS}}$ & \qone & {\small $\mathcal{U} (0, 1)$} & -- \\
    \noalign{\smallskip}
    & $q_{2_\mathrm{TESS}}$ & \qtwo & {\small $\mathcal{U} (0, 1)$} & -- \\
    \noalign{\smallskip}
    
    \midrule
    
    Derived parameters &&&& \\
    \quad Stellar inclination & $i_\star$ & \istar & -- & deg \\
    \noalign{\smallskip}
    \quad Stellar oblateness & $f_\star$ & \fstar & -- & \% \\
    \noalign{\smallskip}
    \quad Stellar rotation period & $P_\star$ & \Pstar & -- & d \\
    \noalign{\smallskip}
    \quad Limb darkening coefficients &&&& \\
    \multirow{2}{*}{\qquad Quadratic coefficients} & $u_{1_\mathrm{TESS}}$ & \uone & -- & -- \\
    \noalign{\smallskip}
    & $u_{2_\mathrm{TESS}}$ & \utwo & -- & -- \\
    \noalign{\smallskip}
    \quad Spin-orbit angle & $\Psi$ & \PSI & -- & deg \\
    \noalign{\smallskip}
    \quad Orbital inclination & $i_p$ & \iplanet & -- & deg \\
    \noalign{\smallskip}
    \quad \boldchange{Planetary radius} & $R_p$ & \RpJup & -- & R$_\text{jup}$ \\
    \noalign{\smallskip}
    \quad \boldchange{Semi-major axis} & $a$ & \asemimajor & -- & AU \\
    \noalign{\smallskip}
    \quad Occultation depth & $F_d/F_\star$ & \FdayFs & -- & ppm \\
    \noalign{\smallskip}
    \quad Nightside flux & $F_n/F_\star$ & \FnFs & -- & ppm \\
    \noalign{\smallskip}
    \quad Phase offset & $\phi_{\text{off}}$ & \phioff & -- & deg \\
    \bottomrule
    \end{tabular}
    \tablefoot{The Gaussian priors with mean $\mu$ and variance $\sigma^2$ are displayed as $\mathcal{N}(\mu, \sigma)$. $\mathcal{U}(a,b)$ shows the uniform prior between $a$ and $b$.}
    \label{tab:fitted_transit_param}
\end{table*}

\newpage

\section{TESS light curves of WASP-189\,b reduced from PDC-SAP and SCALPELS pipeline}

\begin{figure*}[h!]
    \centering
    \includegraphics[width=\textwidth]{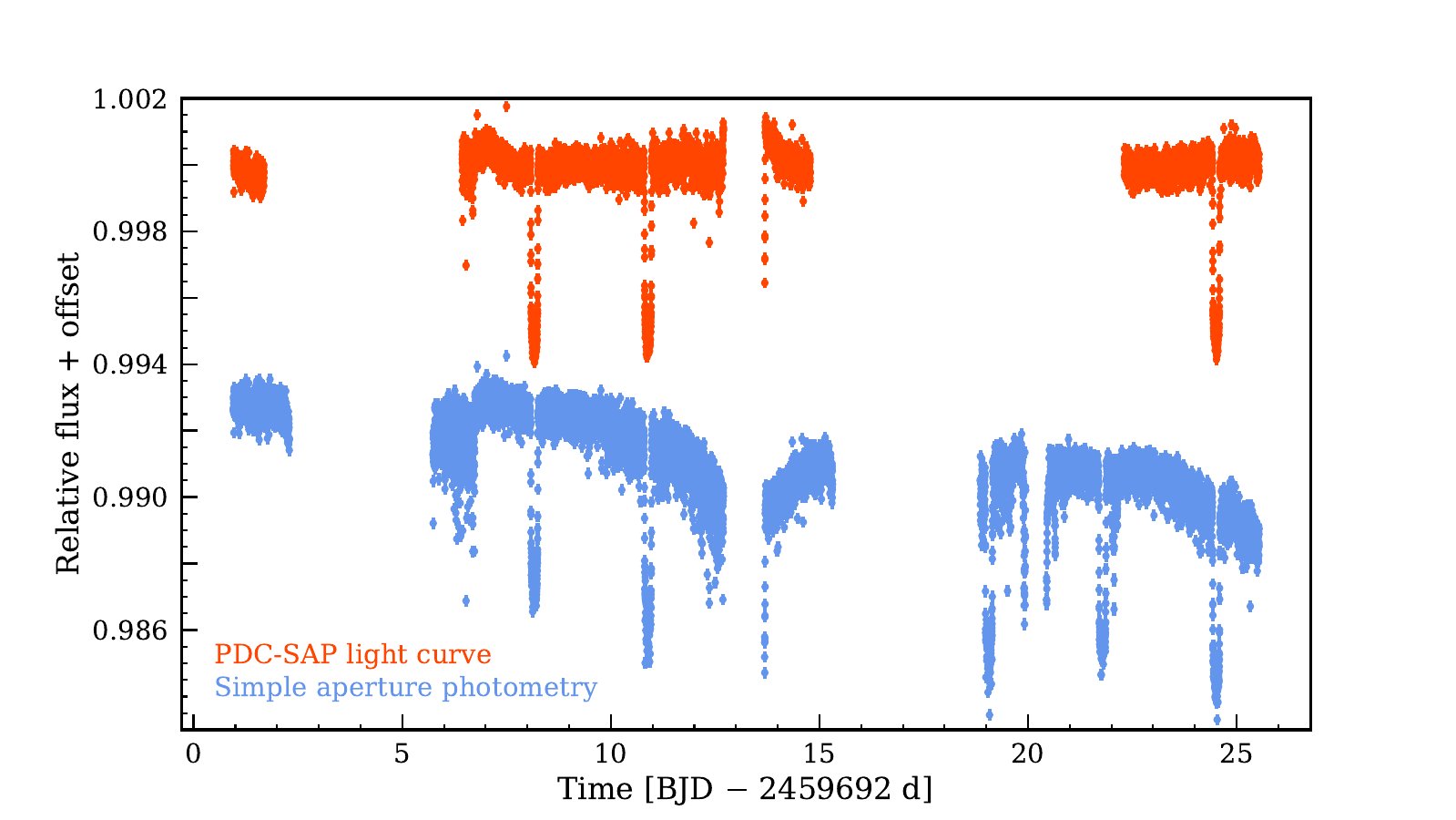}
    \caption{Light curves of WASP-189\,b reduced from two pipelines. The PDC-SAP light curve produced by TESS SPOC is shown in orange, and the light curve extracted using our manual SCALPLES pipeline is shown in blue.}
    \label{fig:pdc_scal_comp}
\end{figure*}

\newpage
\section{GP models}

\begin{figure}[h!]
    \centering
    \includegraphics[width=\textwidth]{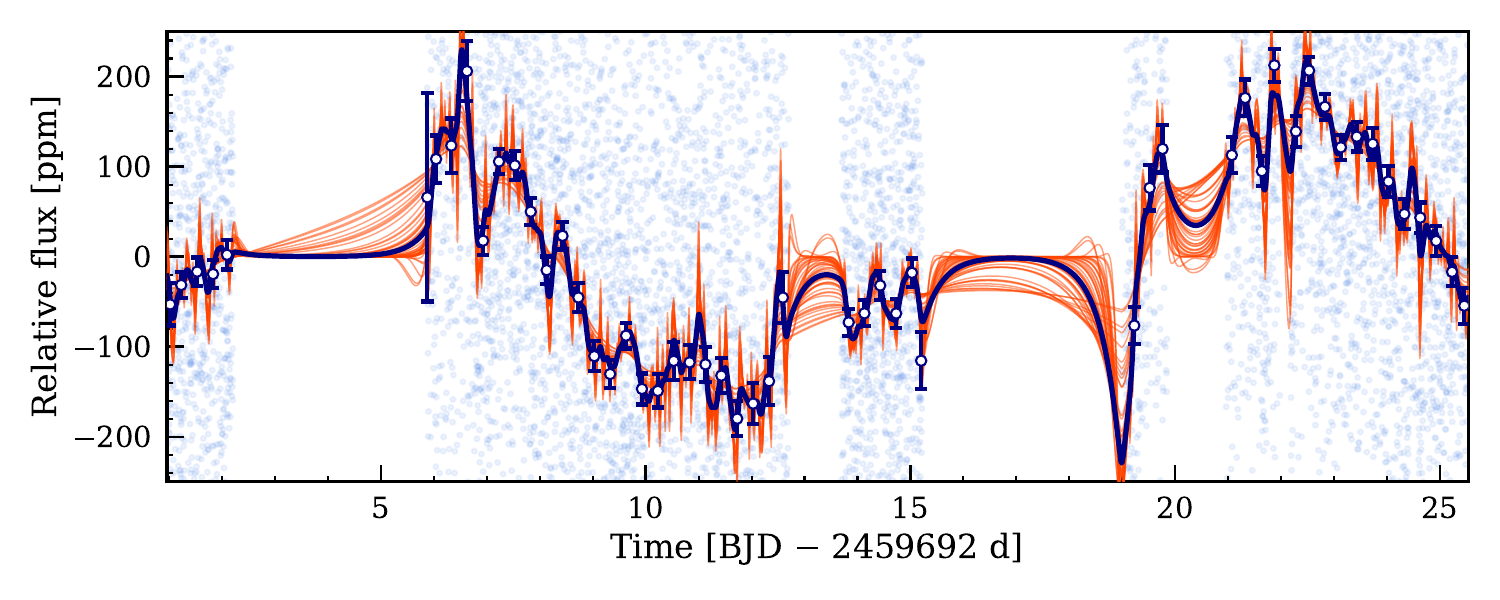}
    \includegraphics[width=\textwidth]{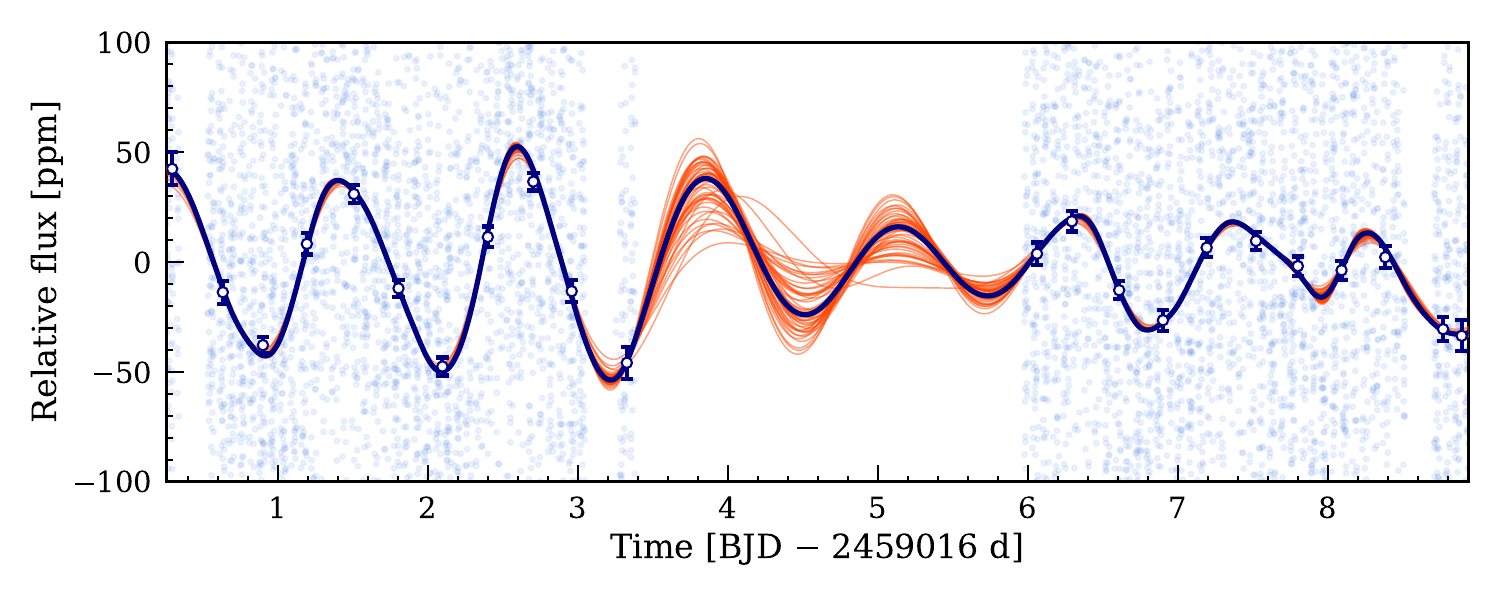}
    \caption{Best-fitted GP models to the TESS (top) and CHEOPS (bottom) data. The light and dark blue points are unbinned and binned data points, respectively. The solid blue and orange lines give the median GP model and models generated from randomly selected posteriors.}
    \label{fig:GP_models}
\end{figure}

\begin{table}[h!]
    \caption{GP parameters used in the light curve analysis}
    \centering
    \begin{tabular}{lcccc}
        \toprule
        \toprule
        Parameters & Symbols & Values (TESS) & Values (CHEOPS) & Units \\
        \midrule
        Amplitude & $\ln{S0}$ & \GPS & {\small $-22.99 ^{+0.48} _{-0.57}$} & -- \\
        \noalign{\smallskip}
        Quality factor & $\ln{Q}$ & \GPQ & {\small $1.33 ^{+0.98} _{-0.73}$} & -- \\
        \noalign{\smallskip}
        Undamped period & P0 & \GPundamped & {\small $1.39 ^{+0.03} _{-0.03}$} & d \\
        \bottomrule
    \end{tabular}
    \label{tab:GP_fitted_params}
\end{table}

\end{appendix}

\end{document}